\documentclass[preprint,12pt]{elsarticle}

\usepackage{amssymb}
\usepackage{amsmath}
\usepackage{booktabs}
\usepackage{textcomp, gensymb}
\usepackage{amsfonts,amssymb}
\usepackage{xcolor}
\usepackage{colortbl}
\usepackage{booktabs}
\usepackage{multirow}
\usepackage{comment}
\usepackage{makecell}
\usepackage{enumitem}
\usepackage{subcaption}
\usepackage{algorithm}
\usepackage{float}
\usepackage{amsmath}
\usepackage{amssymb}
\usepackage{tabularx}
\usepackage{tabulary}
\usepackage{algorithmicx}
\usepackage{algpseudocode}
\usepackage{hyperref} 
\usepackage{lineno}
\usepackage{zref-base}
\usepackage{zref-lastpage}
\usepackage{pifont} 
\providecommand{\cmark}{$\checkmark$}
\providecommand{\xmark}{$\times$}
\definecolor{lightgreen}{RGB}{200,255,210}

\makeatletter
\zref@newprop{linenumber}{\the\c@linenumber}
\zref@addprop{main}{linenumber}
\makeatother

\def\RP#1{{\color{black} {{#1}}}}
\newcommand{\RPblue}{\color{black}\captionsetup{font+={color=black}}}

\begin{document}

\begin{frontmatter}


\author[1]{Rishov Paul}

\author[2]{Sona Ghadimi}

\author[3]{Cristiane C. Singulane}

\author[3]{Amit R. Patel}

\author[2]{Frederick H. Epstein}

\author[1]{Miaomiao Zhang\corref{cor1}}

\cortext[cor1]{Corresponding author. Email: mz8rr@virginia.edu}

\affiliation[1]{organization={Department of Computer Science},
            addressline={University of Virginia},
            country={USA}}

\affiliation[2]{organization={Department of Biomedical Engineering},
            addressline={University of Virginia},
            country={USA}}

\affiliation[3]{organization={Department of Medicine, Division of Cardiology},
            addressline={University of Virginia},
            country={USA}}

\title{GenStrain: Multi-Conditional Generative Diffusion Model To Synthesize High-quality Motion and Strain Maps From Cardiac Magnetic Resonance Images}

\author{} 

\begin{abstract}
This paper presents GenStrain, a novel method for synthesizing high-quality cardiac motion-based strain maps from routinely collected cine cardiac magnetic resonance (CMR) images. \RP{Existing approaches typically rely on image registration to estimate motion (a sequence of displacements) from cine CMR images, but cine CMR presents very sparse trackable intramyocardial features, presenting challenges to estimate intramyocardial motion.} An advanced imaging technique, Displacement Encoding with Stimulated Echoes (DENSE) CMR, is considered as the gold standard for strain assessment, but it requires additional acquisitions, which are not ideal for efficient clinical workflows. 
To address this gap, we introduce GenStrain, a novel framework designed to learn the data distribution of highly accurate cardiac motion captured by DENSE. In particular, we develop a diffusion process in motion space, where spatiotemporal continuity is carefully considered. \RP{To fully leverage both cine CMR images and their associated motion patterns, we introduce a multi-conditional conditioning strategy that integrates anatomical features and motion representations learned by a spatiotemporal registration model to guide the generation process. Experimental results demonstrate that our proposed model, GenStrain, outperforms existing strain analysis methods based on displacement estimates from both traditional registration approaches and state-of-the-art (SOTA) deep learning models. Our model represents a promising step toward more accurate cardiac function assessment and holds great potential for clinical translation}. Our code is publicly available at \emph{https://github.com/Rishov-MIA/GenStrain}.
\end{abstract}

\begin{keyword}
Generative Diffusion Models \sep CMR Image Registration \sep Myocardial Strain Analysis \sep DENSE Strain Imaging
\end{keyword}

\end{frontmatter}

\makeatletter
\ifcsname tabulary*\endcsname\else
\newenvironment{tabulary*}{%
  \let\TY@div\Gscale@div
  \def\Gscale@div##1##2##3{%
    \TY@div{##1}{##2}{##3}%
    \count@\z@ \@tempswafalse
    \loop\ifnum\count@<\TY@count \advance\count@\@ne
      \expandafter\ifx\csname TY@F\the\count@\endcsname\maxdimen
        \@tempswatrue
        \expandafter\xdef\csname TY@\the\count@\endcsname{\the\dimexpr
          \csname TY@\the\count@\endcsname*\number##2/\number##3\relax}%
      \fi
    \repeat
    \if@tempswa \global##3=##2\relax \def##1{1}\fi}%
  \tabulary}{\endtabulary}
\fi
\makeatother
\section{Introduction}
Myocardial strain analysis of cine CMR imaging provides important quantitative biomarkers for evaluating cardiac function and detecting select types of heart disease such as cardiomyopathies, chemotherapy-induced cardiotoxicity, and mechanical dyssynchrony~\cite{scatteia2017strain,rajiah2022myocardial,gulsin2026cardiovascular}. These measurements play a critical role in clinical decision-making by enabling improved diagnosis, risk stratification, and personalized treatment planning for patients with heart disease~\cite{fischer2020feature,xing2021deep,chadalavada2024myocardial}. Various techniques have been developed to characterize cardiac motion and deformation throughout the cardiac cycle to effectively quantify myocardial strain from routinely acquired cine CMR images~\cite{markl2011comprehensive,xing2024lamod,morais2017cardiovascular}. Other efforts relied on CMR tagging, which introduces spatial magnetization patterns into the myocardium that deform with tissue motion, providing physically interpretable measurements and serving as a reference standard for myocardial strain quantification~\cite{ferdian2020fully,shehata2009myocardial,zerhouni1988human}. However, its clinical practicality is limited by the need for dedicated pulse sequences, increased acquisition time, and labor-intensive post-processing~\cite{bucius2020comparison,cao2018comparison}. 

To improve workflow efficiency, CMR feature tracking (FT)~\cite{schuster2016cardiovascular,xu2022state} has been developed as an automated technique that estimates myocardial motion directly from standard cine CMR images. It uses registration-based methods (e.g. optical flow) to track changes in image intensity. With recent advances in deep learning, several studies have leveraged registration-based networks to efficiently predict myocardial deformations and strain directly from standard cine CMR images~\cite{morales2021deepstrain,xing2024lamod,wang2023strainnet}. These approaches have demonstrated significant improvements in spatiotemporal resolution of motion estimation, which provide rapid, automated, and potentially more accurate myocardial strain quantification. In particular, to ensure temporally coherent deformation fields, researchers have incorporated temporal consistency constraints to maintain realistic motion trajectories across cardiac frames~\cite{wu2024tlrn,perperidis2005spatio}. In addition, biomechanical priors have been leveraged to embed knowledge of tissue properties and cardiac mechanics, improving the physical fidelity of the predicted motion~\cite{qin2020biomechanics,qin2023generative}. These works have established deep learning–based approaches as a powerful and flexible framework for automated myocardial motion and strain analysis from standard CMR data with improved clinical accessibility. However, the heavy reliance on image appearance and temporal intensity changes makes the aforementioned approaches sensitive to low tissue contrast, through-plane motion, and complex deformation patterns; hence leading to reduced precision in the subsequent strain estimation~\cite{morais2017cardiovascular,wehner2018comparison}. This challenge is further compounded by the relative scarcity of distinctive, trackable features within the cardiac myocardium.

In parallel, DENSE-CMR has emerged as a gold standard for myocardial strain imaging, as it directly encodes tissue displacement into the MR signal phase, providing high spatial resolution and quantitative measurements~\cite{zhong2010imaging, spottiswoode2007tracking, ghadimi2023improved}. Studies have shown that DENSE achieves superior accuracy and reproducibility in regional and segmental strain compared to FT and related deep learning approaches, and it is frequently used as benchmark for validation~\cite{amzulescu2019myocardial,zhong2010imaging,kim2004myocardial}. Despite these advantages, DENSE is limited by longer acquisition times and restricted availability on commercial scanners. As a result, DENSE remains primarily a research tool rather than a routine clinical technique. \RP{To bridge the gap between the practicality of standard cine CMR and the high accuracy of DENSE, our group has initiated the development of DENSE-guided deep learning networks that use a regression-based approach to predict myocardial displacement directly from routine cine images~\cite{wang2023strainnet}. Although this approach demonstrated the potential of leveraging DENSE supervision to improve strain estimation, it did not fully exploit the rich motion information embedded in the input cine images. To address this issue, a generative diffusion model was developed to learn registration-based motion representations for DENSE reconstruction~\cite{xing2024lamod}. Specifically, the model learned the distribution of motion features derived from image registration and generated these features to further refine the estimated motion. Despite its effectiveness, this approach relies mainly on the quality of image registration and does not explicitly incorporate anatomical information in the reconstruction process.}

In this paper, we present GenStrain, a multiconditional generative diffusion model developed to synthesize high-quality, anatomically and temporally coherent cardiac motion/deformation directly from standard cine CMR images. \RP{In contrast to previous closely related approaches~\cite{wang2023strainnet,xing2024lamod}, GenStrain is the first to directly learn the distribution of high-quality cardiac motion captured by DENSE, conditioned on multi-modal features that integrates anatomical information and motion representations derived from the input images.} Our contributions are summarized as follows:
\begin{enumerate}[label=(\roman*).]
    \item \RP{Develop a novel generative diffusion model, GenStrain, that directly learns the distribution of high-quality DENSE motion; hence improving the fidelity of synthesized strain maps across diverse disease conditions.}  
    \item Introduce a multi-conditional mechanism that integrates anatomical information and motion dynamics learned by a registration network. A spatiotemporal cross-attention module is particularly designed to effectively fuse motion and intensity information to guide the generation process toward coherent and realistic motion.
    \item Demonstrate the utility of GenStrain on CMR images collected from multiple clinical sites. Experimental results show substantial improvements in both motion prediction and strain quantification when compared to existing methods.
\end{enumerate}

\RP{Our work highlights the potential of GenStrain as a clinically practical solution for improving myocardial motion and strain analysis using routine CMR, achieving performance comparable to strain-dedicated acquisitions. Additionally, our GenStrain can be readily integrated into existing post-processing workflows, which may facilitate a more flexible and broader clinical adoption.}

\section{Background: Registration-based Motion Estimation} 
\label{sec:background}
In this section, we will review a general mathematical framework for motion estimation based on deformable image registration~\cite{wu2024tlrn,chandrashekara2003construction,pan2021efficient}. The goal is to estimate displacement fields that describe myocardial motion over time by aligning a temporal image sequence to a reference frame.

Let $\Omega \subset \mathbb{R}^d$ denote a $d$-dimensional image domain. Given a reference image $\mathcal{I}_0:\Omega \rightarrow \mathbb{R}$ and a sequence of moving images $\{\mathcal{I}_t\}_{t=1}^{T}, t \in \{1, \ldots, T\}$, the objective is to estimate a sequence of spatial transformations, $\{\phi_t\}_{t=1}^{T} : \Omega \rightarrow \Omega$, that map spatial locations from the reference frame to the target frame at time $t$. The corresponding displacement field is defined as $\eta_t = \phi_t - \mathrm{Id}$, where $\mathrm{Id}$ denotes an identity transformation. 

A classical deformable registration~\cite{sotiras2013deformable,beg2005computing,arsigny2006log} formulates motion estimation as an energy minimization problem over a sequence of transformations $\{\phi_t\}$ as
\begin{equation}
\text{min} \, \mathcal{L}(\phi_t) = \sum_{t=1}^{T}
\mathrm{Dist}\!\left(\mathcal{I}_0 \circ \phi_t^{-1},\, \mathcal{I}_t \right)
+ \lambda \, \mathrm{Reg}(\phi_t),
\label{eq:temporal_registration_energy}
\end{equation}
where $\circ$ denotes an interpolation operator, $\mathrm{Dist}(\cdot,\cdot)$ measures image dissimilarity, and $\mathrm{Reg}(\cdot)$ is a regularization term that enforces the smoothness of the transformation field. The parameter $\lambda>0$ controls the trade-off between image matching fidelity and deformation regularity. 
The regularization term typically penalizes the tangent vector of the transformation $\phi_t$, which is also known as velocity fields and denoted by $v_t$~\cite{beg2005computing,arsigny2006log,miller2004computational}. Various choices of regularizations, such as large deformation diffeomorphic metric mapping (LDDMM)~\cite{beg2005computing} and stationary velocity field (SVF)~\cite{arsigny2006log}, enforce diffeomorphic constraints (a.k.a. one-to-one smooth and inverse smooth mapping) to ensure anatomically plausible transformations. 

\paragraph*{\bf Parameterization of deformations} In the LDDMM framework~\cite{beg2005computing,miller2004computational}, 
each temporal deformation, $\phi_t$, is modeled as a flow of smooth and invertible smooth transformations (also known as diffeomorphisms), parameterized by the velocity field $v_t(\tau)$, where $\tau \in [0,1]$ indicating the flow sequence. The deformation $\phi_t$ evolves according to the flow equation,
\begin{equation}
\frac{d\phi_t (\tau)}{d \tau}  = v_t (\tau) \circ \phi_t (\tau),
\quad \phi_t(0) = \mathrm{Id}.
\label{eq:lddmm_flow}
\end{equation}
The optimal deformation, $\phi_t (\tau=1)$,  can be uniquely determined by integrating the Euler–Poincaré differential equation (EPDiff)~\cite{arnold1966,miller2006geodesic} along a geodesic path with a given initial condition~\cite{vialard2012}. 

Another commonly used regularization, the SVF framework~\cite{arsigny2006log,ashburner2007fast}, simplifies the aforementioned LDDMM by assuming the velocity field is constant over time, i.e., $v_t (\tau) = v_t (0)$ for all $\tau \in [0,1]$ in Eq.~\eqref{eq:lddmm_flow}. 
The resulting diffeomorphic transformation can then be given by the group exponential map, i.e., 
\begin{equation}
\label{eq:svf}
\phi_t(\tau) = \mathrm{Exp}(v_t(0)).
\end{equation}
In practice, this exponential map is efficiently computed via the scaling-and-squaring method. More details can be found in~\cite{arsigny2006log,ashburner2007fast}. 

In this paper, we adopt commonly used sum-of-squared intensity differences (SSD) as the similarity metric~\cite{wu2024tlrn,rueckert2006diffeomorphic}, and SVF as the regularity term. For notation simplicity, we omit the flow index $\tau$ in the following sections. Consequently, the objective function for temporal registration in Eq.~\eqref{eq:temporal_registration_energy} can be formulated as
\begin{equation}
\label{eq:tlrn}
    \text{min} \, \mathcal{L}(v_t) = \sum_{t=1}^{T}
\| \mathcal{I}_0 \circ \phi_t^{-1}(v_t) - \mathcal{I}_t \|^2_2
+ \lambda \| \nabla v_t \|_2^2,  \, \, s.t. Eq.~\eqref{eq:svf}. 
\end{equation}

\subsection{Myocardial strain computation from motion fields} Once displacement fields are estimated, myocardial strain can be computed from the local deformation gradient tensor~\cite{xing2024lamod,kindberg2012myocardial}, i.e., $F_t = \nabla \phi_t = \mathrm{Id} + \nabla \eta_t$, where $\nabla \eta_t$ is the spatial Jacobian of the displacement field. For finite deformation analysis, myocardial strain, $S_t:\Omega \rightarrow \mathbb{R}$, is commonly quantified using the Green-Lagrange strain tensor~\cite{spottiswoode2007tracking,smiseth2016myocardial} at each pixel/voxel, i.e.,
\begin{equation}
\label{eq:strain}
S_t = \frac{1}{2}\big(F_t^{T} F_t - \mathrm{Id} \big).
\end{equation}
This paper focuses on the left ventricular (LV) myocardial circumferential strain, which measures the strain component along the myocardial wall. In clinical practice, the LV myocardium is often divided into a variable number of segments for regional analysis~\cite{american2002standardized}. In this study, we divide each short-axis slice into six segments at the basal and mid-ventricular levels, and four segments at the apical level~\cite{american2002standardized}. Additional details will be provided in the experimental section.

\section{Our Method: GenStrain}
\begin{figure}[htbp]
    \centering
    \includegraphics[width=\textwidth]{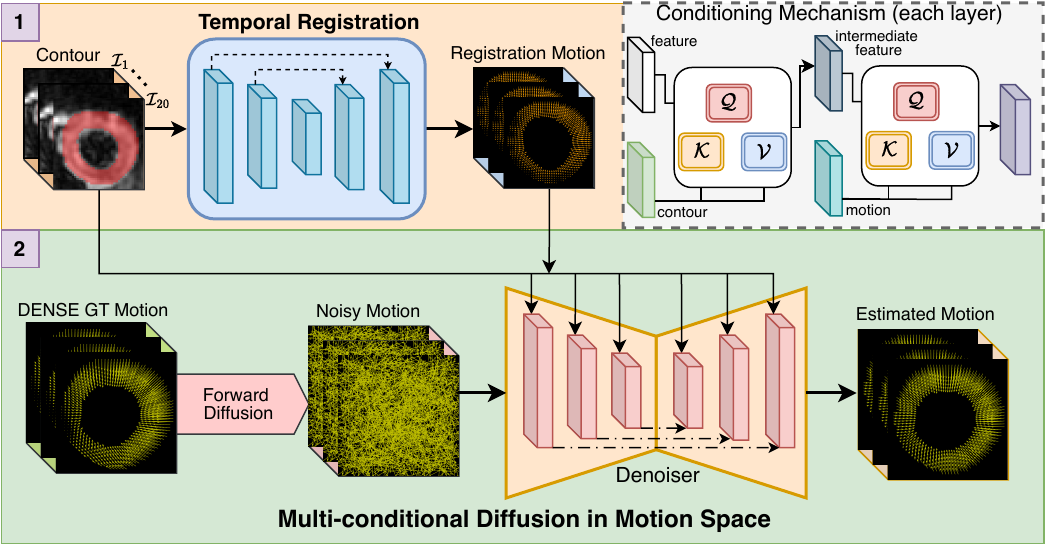}
    \caption{Our proposed framework, GenStrain, consists of two main components: (i) an unsupervised temporal latent registration network to extract spatiotemporal motion from CMR images; and (ii) a multi-conditional diffusion model in motion space to promote high-fidelity displacement fields.}
    \label{fig:architecture}
\end{figure}

In this section, we introduce a multi-conditional generative diffusion model, GenStrain, in a spatiotemporal displacement space to directly learn the distribution of high-quality DENSE displacement fields conditioned on both CMR images and registration motion. Our framework includes two key modules: (i) a time-series registration network~\cite{wu2024tlrn} that predicts spatiotemporal displacement/motion from input CMR image sequences; and (ii) a generative diffusion model that learns the high-fidelity motion distribution from DENSE displacement fields, conditioned on joint image and motion features. Once trained, our model, GenStrain, will generate high-quality displacement fields for subsequent analysis of myocardial strain given routinely collected cine CMRs. An overview of GenStrain is illustrated in Fig.~\ref{fig:architecture}. 

\subsection{Motion learning from spatiotemporal registration network}
\label{subsec:motionlearning}
Given a training dataset of $N$ cardiac image sequences, $\{\mathcal{I}^i\}_{i=1}^N, i \in \{1, \ldots, N\}$, where each sequence $\mathcal{I}^i$ contains $T+1$ temporal frames, i.e., $\mathcal{I}^i = \{ \mathcal{I}^{i}_{t} \}_{t=0}^T$. Our objective is to predict a sequence of deformation fields, $\{\phi_t^i\}$, associated with the reference image and following images as $\{ \mathcal{I}^i_0, \mathcal{I}^i_t \}_{t=1}^{T}$. Analogous to~\cite{wu2024tlrn}, we employ a spatiotemporal registration network to predict the velocity fields $v_t(\zeta)$, parameterized by $\zeta$. Following a similar principle of Eq.~\eqref{eq:tlrn}, we formulate the corresponding network loss function as
\begin{equation}
\label{eq:reg_loss}
\mathcal{L}(\zeta)
=
\sum_{i,t=1}^{N,T}
\left\|
\mathcal{I}_0^i \circ \phi_t^i(v_t^i(\zeta)) - \mathcal{I}_t^i
\right\|_2^2
+
\sum_{t=1}^{T}
\lambda
\left\|
\nabla v_t^i(\zeta)
\right\|^2
+
\sum_{t=2}^{T}
\beta
\left\|
\Delta v_t^i(\zeta)
\right\|^2,
\end{equation}
where $\Delta v^i_t(\zeta) \triangleq v_t^i(\zeta) - v_{t-1}^i(\zeta)$, representing a temporal regularization to promote temporal continuity, and $\beta$ is a weighting parameter. 

Given transformation fields $\{\phi_t^i\}$ computed by the predicted velocity fields $\{v_t^i\}$, we obtain the associated displacement fields as $\boldsymbol{\eta}_t^i = \phi_t^i - \mathrm{Id}$. Note that, in this paper, we adopt a U-Net backbone with a latent residual network to explicitly model temporal correlations in the registration framework~\cite{wu2024tlrn}. Our proposed methodology is architecture-agnostic, and alternative backbone designs~\cite{balakrishnan2019voxelmorph,chen2022transmorph} can be readily incorporated without modification to the overall formulation.

\subsection{Multi-Conditional Diffusion Model In Motion Space}
We develop a conditional diffusion process in the {\em vectorized motion space} that effectively learns the underlying spatiotemporal distribution of high-fidelity DENSE displacements; hence enabling the generation of realistic displacement trajectories given any input CMR image sequences.

Let $\mathbf{u}_0 \in \mathbb{R}^{d \times T \times H \times W}$ denote the high-fidelity DENSE displacements over an entire cardiac cycle (e.g., $d=2$ for in-plane motion), where $H$ and $W$ represent the spatial resolution of each frame. Consider the data $\mathbf{u}_0 \sim q(\mathbf{u}_0)$, our model aims to learn a parameterized conditional distribution of displacement trajectories, $p_\theta(\mathbf{u}_0 \mid \mathcal{C})$, where $\theta$ denotes model parameters, and \(\mathcal{C}\) represents a set of conditioning signals that encode prior information such as anatomical structure, image appearance, or motion dynamics.

Following a similar principle of~\cite{ho2020denoising}, the conditional distribution can be expressed by marginalizing a diffusion trajectory in motion space, denoted as $\{\mathbf{u}_s\}_{s=1}^S$, where each $\mathbf{u}_s \in \mathbb{R}^{d \times T \times H \times W}$ represents the displacement field at diffusion step $s$, and $S$ denotes the total number of diffusion steps. We have
\begin{align}
    p_\theta(\mathbf{u}_0 \mid \mathcal{C}) &= \int p_\theta(\mathbf{u}_{0:S} \mid \mathcal{C}) \, d \, \mathbf{u}_{1:S}, \notag \\
    &= \int p(\mathbf{u}_S)\prod_{s=1}^{S} p_\theta(\mathbf{u}_{s-1} \mid \mathbf{u}_s, \mathcal{C})  d \, \mathbf{u}_{1:S}.
\label{eq:marginal}
\end{align}

We are now ready to formulate a conditional diffusion model to learn the reverse transition distribution $ p_\theta(\mathbf{u}_{s-1} \mid \mathbf{u}_s, \mathcal{C}) $ in Eq.~\eqref{eq:marginal}. Similar to~\cite{ho2020denoising}, we define a forward diffusion process that progressively perturbs the data $\mathbf{u}_0$ by injecting Gaussian noise over $S$ diffusion steps. This transforms the structured displacement field into a tractable isotropic Gaussian distribution $\epsilon \sim  \mathcal{N}(0, \mathbf{I})$. The reverse process is then parameterized to approximate this diffusion trajectory in reverse time via a sequence of conditional distributions. Under this formulation, our model effectively learns the complex distribution of high-quality myocardial DENSE motion, while incorporating both structural and dynamic motion conditioning. Details of the forward and reverse processes are described below.

\paragraph*{\bf Forward Process} We define the forward diffusion process in the motion space by progressively corrupting the high-quality DENSE displacement field, $\mathbf{u}_0$, by a variance schedule $\{\beta_s\}_{s=1}^S$, where $\beta_s \in (0,1)$. Note that this process is independent of the conditioning variables and transforms structured data into noise via a Markov chain
$q(\mathbf{u}_{1:S} \mid \mathbf{u}_0) = \prod_{s=1}^{S} q(\mathbf{u}_s \mid \mathbf{u}_{s-1}),$
where each transition is defined as
\begin{equation*}
    q(\mathbf{u}_s \mid \mathbf{u}_{s-1}) = \mathcal{N}\big(\mathbf{u}_s; \sqrt{1 - \beta_s}\,\mathbf{u}_{s-1}, \, \beta_s \mathbf{I}\big).
\end{equation*}

Let $\alpha_s = 1 - \beta_s$ and $\bar{\alpha}_s = \prod_{j=1}^s \alpha_j$. Instead of performing the forward process step-by-step, we can directly sample $\mathbf{u}_s$ at an arbitrary timestep $s$ using the closed-form solution~\cite{ho2020denoising}, i.e.,
\begin{equation}
\label{eq:forward}
q(\mathbf{u}_s \mid \mathbf{u}_0) = \mathcal{N}\big(\mathbf{u}_s; \sqrt{\bar{\alpha}_s}\,\mathbf{u}_0, \, (1 - \bar{\alpha}_s)\mathbf{I}\big),
\end{equation}
which can be equivalently written as $\mathbf{u}_s = \sqrt{\bar{\alpha}_s}\,\mathbf{u}_0 + \sqrt{1 - \bar{\alpha}_s}\,\boldsymbol{\epsilon}$.

\paragraph*{\bf Conditional Reverse Process} Our reverse diffusion process is designed to reconstruct the original displacement field, $\mathbf{u}_0$, from pure Gaussian noise through an encoder-decoder based denoiser network $\boldsymbol{\epsilon}_\theta$ parametrized by $\theta$.
At each timestep $s$, our model GenStrain approximates the reverse transition by a conditional Gaussian distribution formulated as
\begin{equation}
p_\theta(\mathbf{u}_{s-1}|\mathbf{u}_s, \mathcal{C}) = \mathcal{N}(\mathbf{u}_{s-1};\, \boldsymbol{\mu}_\theta(\mathbf{u}_s, s, \mathcal{C}),\, \Sigma_s^2 \mathbf{I}),
\end{equation}
where $\boldsymbol{\mu}_\theta$ is the learned mean and \(\Sigma_s^2\) is a predefined variance schedule. 

Instead of directly parameterizing the reverse-transition mean $\boldsymbol{\mu}_\theta$, we adopt the common reparameterization strategy of predicting the additive Gaussian noise $\boldsymbol{\epsilon}$ introduced in the forward diffusion process~\cite{ho2020denoising}. Given $\mathbf{u}_s$ (defined in Eq.~\eqref{eq:forward}), the reverse mean $\boldsymbol{\mu}_\theta$ can then be computed from the predicted noise $\boldsymbol{\epsilon}_\theta$ as
\begin{equation}
\label{eq:rev_mean}
\boldsymbol{\mu}_\theta(\mathbf{u}_s, s, \mathcal{C}) =
\frac{1}{\sqrt{\alpha_s}}
\left(
\mathbf{u}_s -
\frac{1-\alpha_s}{\sqrt{1-\bar{\alpha}_s}}
\boldsymbol{\epsilon}_\theta(\mathbf{u}_s, s, \mathcal{C})
\right).
\end{equation}

Learning the conditional reverse process therefore simplifies to training the noise estimator, $\boldsymbol{\epsilon}_\theta$. The training objective and inference-time sampling procedure are described below.

\paragraph*{\bf Multi-conditioning Mechanism} Our reverse diffusion process is conditioned on a number of $M$ sources, i.e., $\mathcal{C} \triangleq \{ \mathcal{C}^{(i)} \}_{i=1}^M$, that integrate both image and motion features. In practice, these conditioning are incorporated into the denoising network $\boldsymbol{\epsilon}_\theta$ via a multi-head cross-attention mechanism~\cite{vaswani2017attention}, applied at multiple resolution levels. Specifically, at each layer \(l \in \{1, \dots, L\}\), we extract level-specific conditioning features \(\mathcal{C}^{(i)}_l\) and iteratively update the latent representation \(\mathcal{U}_l\) through a sequence of operations. Each update is performed via a \emph{spatiotemporal cross-attention block}, which consists of a spatial attention module followed by a temporal attention module. Mathematically, we formulate this as
\begin{equation}
\mathcal{U}_l \leftarrow \mathcal{U}_l 
+ \mathrm{Attn}_{\mathrm{temporal}}\!\Big(
    \mathcal{U}_l 
    + \mathrm{Attn}_{\mathrm{spatial}}\big(\mathcal{U}_l,\, \mathcal{C}^{(i)}_l\big),\;
    \mathcal{C}^{(i)}_l
\Big). 
\end{equation}
Here, each attention module follows the standard scaled dot-product formulation, where queries are derived from the latent representation and keys/values from the conditioning as
\begin{equation}
\mathrm{Attn}\!\big(\mathcal{U},\, \mathcal{C}^{(i)}_l\big) 
= \mathrm{softmax}\!\left(\frac{\mathbf{Q}\mathbf{K}^\top}{\sqrt{d_k}}\right)\mathbf{V},
\end{equation}
with $\mathbf{Q} = \mathbf{W}_Q \mathcal{U}, 
\quad 
\mathbf{K} = \mathbf{W}_K \mathcal{C}^{(i)}_l, 
\quad 
\mathbf{V} = \mathbf{W}_V \mathcal{C}^{(i)}_l,$ where \(\mathbf{W}_Q, \mathbf{W}_K, \mathbf{W}_V\) are learnable projection matrices and \(d_k\) is the key dimension. This hierarchical conditioning mechanism enables the denoising network to be adaptively modulated by multiple priors at each layer of the latent feature representation. Moreover, our formulation naturally generalizes to an arbitrary number of conditioning signals, as each additional source is incorporated through the same sequential cross-attention updates.

\RP{In our experiments, we specifically incorporate two conditioning signals ($M = 2$) to promote anatomical consistency and motion fidelity: the left ventricle (LV) segmentation mask derived from images, $\boldsymbol{\omega}$, and the motion features, $\boldsymbol{\eta}$, learned by our temporal registration module.}

\subsection{Training Objective \& Inference}
\label{para:objective}
Following the principle of~\cite{ho2022video}, our model GenStrain is trained by optimizing the following objective function
\begin{equation}
\label{eq:diffusion_loss}
\min_{\theta}\ \mathcal{L}(\theta) = \mathbb{E}_{\mathbf{u}_0, s, \boldsymbol{\epsilon}, \boldsymbol{\omega}, \boldsymbol{\eta}} \left[ \left\| \boldsymbol{\epsilon} - \boldsymbol{\epsilon}_\theta(\mathbf{u}_s, s, \boldsymbol{\omega}, \boldsymbol{\eta}) \right\|^2 \right].
\end{equation}
It is worth noting that this training objective is closely connected to denoising score matching~\cite{song2020score}. From the forward process in Eq.~\eqref{eq:forward}, the noisy sample \(\mathbf{u}_s\) can be written as a closed-form score matching formulation $\nabla_{\mathbf{u}_s}\!\log q(\mathbf{u}_s \mid \mathbf{u}_0)
= -\,\frac{\boldsymbol{\epsilon}}{\sqrt{1 - \bar{\alpha}_s}}$, which shows that the injected noise \(\boldsymbol{\epsilon}\) is proportional to the score function. Therefore, training the denoising network $\boldsymbol{\epsilon}_\theta$ using the objective in Eq.~\eqref{eq:diffusion_loss} is equivalent to learning a scaled estimate of the conditional score function
\begin{equation*}
\boldsymbol{\epsilon}_\theta(\mathbf{u}_s, s, \boldsymbol{\omega}, \boldsymbol{\eta})
\approx -\sqrt{1 - \bar{\alpha}_s}\;\nabla_{\mathbf{u}_s}\!\log p(\mathbf{u}_s \mid \boldsymbol{\omega}, \boldsymbol{\eta}).
\end{equation*}

During inference, we use denoising diffusion implicit model~(DDIM) sampling~\cite{song2020denoising} to generate displacement fields in $K \ll S$ steps, rather than traversing all $S$ diffusion steps. Specifically, we select a subsequence of timesteps, $1 \leq \tau_1 < \cdots < \tau_K = S-1$, uniformly spaced in log signal-to-noise rather than in the original timestep indices, to distribute the denoising steps more evenly across noise levels~\cite{lu2022dpm}. Starting from $\mathbf{u}_{\tau_K} \sim \mathcal{N}(\mathbf{0},\mathbf{I})$, we iteratively denoise the sample for $k=K,\ldots,1$. At each step, the noise estimator predicts $\hat{\boldsymbol{\epsilon}} = \boldsymbol{\epsilon}_\theta(\mathbf{u}_{\tau_k}, \tau_k, \boldsymbol{\omega}, \boldsymbol{\eta})$, from which the gradually denoised displacement field is estimated as $\hat{\mathbf{u}}_0 = (\mathbf{u}_{\tau_k} - \sqrt{1-\bar{\alpha}_{\tau_k}}\,\hat{\boldsymbol{\epsilon}})/\sqrt{\bar{\alpha}_{\tau_k}}$. The sample is then updated using the DDIM transition from timestep $\tau_k$ to $\tau_{k-1}$ as
\begin{equation}
\label{eq:ddim}
\mathbf{u}_{\tau_{k-1}} = \sqrt{\bar{\alpha}_{\tau_{k-1}}}\,\hat{\mathbf{u}}_0
+ \sqrt{1-\bar{\alpha}_{\tau_{k-1}} - \sigma_{\tau_k}^2}\;\hat{\boldsymbol{\epsilon}}
+ \sigma_{\tau_k}\mathbf{z}, \qquad \mathbf{z} \sim \mathcal{N}(\mathbf{0}, \mathbf{I}),
\end{equation}
where $\sigma_{\tau_k} = \gamma \sqrt{(1-\bar{\alpha}_{\tau_{k-1}})/(1-\bar{\alpha}_{\tau_k})}\,\sqrt{1-\bar{\alpha}_{\tau_k}/\bar{\alpha}_{\tau_{k-1}}}$, with $\gamma \in [0,1]$ controls the stochasticity, ranging from deterministic DDIM ($\gamma = 0$) to stochastic DDPM ($\gamma = 1$). 

The training and inference are summarized in Alg.~\ref{alg:train} and Alg.~\ref{alg:inference}.

\begin{algorithm}[h]
\caption{GenStrain Training}
\label{alg:train}
\begin{algorithmic}[1]
\Statex \textbf{Stage 1: Motion Learning from Registration Network}
\Require CMR-image dataset $\{\mathcal{I}_t\}_{t=0}^T$
\Repeat
    \State Predict velocity fields $\{v_t(\zeta)\}_{t=1}^T$
    \State Compute deformation fields $\{\phi_t\}$ from $\{v_t(\zeta)\}$ using Eq.~\ref{eq:svf}
    \State Optimize spatiotemporal registration loss in Eq.~\ref{eq:reg_loss}
\Until converged
\State Obtain registration motions $\{\boldsymbol{\eta}_t\} = \{\phi_t\} - \mathrm{Id}$

\Statex \textbf{Stage 2: Multi-Conditional Diffusion Model in Motion Space}
\Require Displacement fields $\mathbf{u}_0 \in \mathbb{R}^{d \times T \times H \times W}$, noise schedule $\{\beta_s\}_{s=1}^S$
\Require LV segmentation mask $\boldsymbol{\omega}$, motion features $\boldsymbol{\eta}$

\State Precompute $\alpha_s = 1-\beta_s$, $\bar\alpha_s = \prod_{i=1}^s \alpha_i$
\Repeat
    \State Sample $(\mathbf{u}_0, \boldsymbol{\omega}, \boldsymbol{\eta})$ from training data
    \State Sample diffusion timestep $s \sim \{1,\dots,S\}$
    \State Sample $\boldsymbol{\epsilon} \sim \mathcal{N}(0,\mathbf{I})$
    \State Forward process using Eq.~\ref{eq:forward}
    \State Optimize $\mathcal{L}(\theta)$ using Eq.~\ref{eq:diffusion_loss}
\Until converged
\end{algorithmic}
\end{algorithm}

\begin{algorithm}[!h]
\color{black}
\caption{GenStrain Inference}
\label{alg:inference}
\begin{algorithmic}[1]
\Require Timesteps $\{\tau_1  \dots \tau_K\}$, with $\{\bar\alpha_{\tau_k}\}_{k=1}^K$
\Require Conditioning $\boldsymbol{\omega}$ and $\boldsymbol{\eta}$ (predicted from Stage 1)
\State $\mathbf{u}_{\tau_K} \sim \mathcal{N}(\mathbf{0},\mathbf{I})$
\For{$k = K, \dots, 1$}
    \State $\hat{\boldsymbol\epsilon} = \boldsymbol{\epsilon}_\theta(\mathbf{u}_{\tau_k}, \tau_k, \boldsymbol{\omega}, \boldsymbol{\eta})$
    \State Sample $\mathbf{z} \sim \mathcal{N}(\mathbf{0},\mathbf{I})$
    \State Compute $\mathbf{u}_{\tau_{k-1}}$ using Eq.~\ref{eq:ddim}
\EndFor
\State \Return $\mathbf{u}_{\tau_0}$
\end{algorithmic}
\end{algorithm}

\section{Experiments}
\label{sec:experiments}
We evaluate the effectiveness of GenStrain on datasets that include both routinely collected cine CMR and high-quality DENSE CMRs. \RP{We compare its performance against baselines from three main categories: (i) deterministic regression-based models, StrainNet~\cite{wang2023strainnet} and TransUNet~\cite{chen2021transunet}; (ii) publicly available registration-based strain prediction networks, DeepStrain~\cite{morales2021deepstrain} and TLRN~\cite{wu2024tlrn}; and (iii) state-of-the-art generative approaches that are adapted to motion space, flow-matching~\cite{tong2023improving}, ControlNet~\cite{zhang2023adding}, and LaMoD~\cite{xing2024lamod}}.  

\subsection{Dataset}
Our dataset consists of two different CMR modalities: (i) widely available, routinely acquired left-ventricular (LV) short-axis cine CMR images, and (ii) less accessible, high-quality DENSE-CMR acquisitions collected from eight clinical centers with standard harmonization~\cite{xing2024lamod,wang2023strainnet}. \RP{All LV segmentations are manually annotated by clinicians.}

\paragraph*{\bf DENSE CMRs} The DENSE-CMR provides both magnitude images and high-quality displacement field (motion) measurements that serve as training data and a reference standard for subsequent strain evaluation. The DENSE dataset comprises $1,200$ multiphase 2D DENSE short-axis slices from $284$ subjects, including $124$ healthy volunteers and $160$ patients with heart disease~\cite{xing2024lamod,wang2023strainnet}. \RP{DENSE acquisitions have an average in-plane spatial resolution of $2.41 \times 2.41$\,mm$^2$/pixel and a temporal resolution of 15--30\,ms.} The dataset was split into $1,068$ training samples and $132$ testing samples. \RP{Please note that the partition was performed strictly at the subject level, where all slices belonging to a given subject (basal, mid-ventricular, and apical) were assigned exclusively to either the training or the testing set. This precluded leakage between correlated slices of the same heart so that the reported metrics reflect generalization to unseen subjects.}

\paragraph*{\bf Standard cine CMRs} We also evaluate our method on $132$ cine CMRs that are paired with aforementioned held-out DENSE testing dataset collected from 76 subjects (24 healthy volunteers vs. 52 disease patients). \RP{cine CMR acquisitions have an average in-plane spatial resolution of $1.68 \times 1.68$\,mm$^2$/pixel and a temporal resolution of 23--40\,ms.} Since cine CMRs don't provide direct measurements of myocardial motion, each cine sequence is matched to a corresponding DENSE acquisition from the same subject at approximately the same anatomical slice scan location (within $\pm$ 4mm). This spatial correspondence allows the DENSE displacements field to serve as the ground truth motion for the cine CMR videos. 

\paragraph*{\bf Data Preprocessing \& Implementation Details} Prior to training and testing, all CMR data are cropped to a $48^2\times20$ LV region of interest. \RP{Each cardiac cycle is temporally aligned to uniformly spaced phases that cover the entire cardiac cycle. However, we retain the original spatial resolution of each DENSE acquisition to avoid potential smoothing and bias in the displacement fields introduced by spatial interpolation.}

\RP{The training of GenStrain consists of two stages. In the first stage, we train the spatiotemporal registration network (Sec.~\ref{subsec:motionlearning}) using LV segmentations from both DENSE and cine CMRs. This enables the network to learn robust motion representations across variations in myocardial shape, wall thickness, and spatial resolution.} In the second stage, we train our multi-conditional diffusion model solely using the DENSE data. All models in our experiments are trained on an NVIDIA A100 GPU with 80~GB of memory using the Adam optimizer~\cite{Kingma2014AdamAM}. The hyperparameters were selected via grid search during training~\cite{bergstra2012random}, resulting in a batch size of $20$ and a learning rate of $1\times 10^{-5}$. For a fair comparison, all baselines are trained to convergence. \RP{All diffusion-based models used DDIM sampling~\cite{song2020denoising} with $K = 50$ log-SNR-spaced steps and stochasticity parameter $\gamma = 0.8$.}

\subsection{Experimental Design \& Evaluation}
We evaluate all models from multiple perspectives, including: (i) the accuracy of predicted time sequence of displacement fields, (ii) the resulting strain errors relative to reference DENSE data, and (iii) statistical analysis of strain agreement between the predicted models and DENSE measurements. 

\RP{ \paragraph*{\bf Evaluation Metrics} We first evaluate the accuracy of the predicted displacement fields against the DENSE ground truth using the normalized pixel-wise end-point error (nEPE), computed based on the EPE definition in~\cite{xing2024lamod,wang2023strainnet} and normalized with respect to the ground-truth DENSE. Note that we report nEPE metrics exclusively on DENSE CMR data, where the contours and displacement fields share the same spatial coordinate system. Direct displacement comparisons on cine CMR data are not reliable as cine and DENSE acquisitions are not necessarily spatially or temporally aligned.}

\RP{We then validate the accuracy of myocardial circumferential strain based on predicted motion / displacement trajectories using Eq.~\eqref{eq:strain} for both DENSE and cine data.} 
We compare the mean absolute error (MAE) under four evaluation settings: (i) whole-slice strain averaged across cardiac phases, (ii) segmental strain averaged across cardiac phases, (iii) whole-slice end-systolic (ES) strain, and (iv) segmental ES strain~\cite{xing2024lamod,morales2021deepstrain,wang2023strainnet}. Specifically, for segmental strain analysis, we follow the standard clinical convention~\cite{amzulescu2019myocardial} to partition the LV myocardium into six segments (anteroseptal, inferoseptal, inferior, inferolateral, anterolateral, and anterior) for basal and mid-ventricular slices, and four segments (septal, inferior, lateral, and anterior) for apical slices. Such a partition begins at the anterior right ventricular insertion point and proceeds counterclockwise. The segmental strain is evaluated both across all temporal frames and specifically at the ES frame (with a minimal value). The whole-slice strain is obtained by averaging the segmental strain values prior to error computation. All predicted strain measurements are compared against reference strain values derived from the paired high-quality DENSE ground-truth displacement fields. \RP{Note that myocardial circumferential strain is expressed as a percentage, with typical peak-systolic magnitudes reaching approximately $25\%$, rather than the $100\%$ maximum commonly associated with conventional percentage-based metrics~\cite{yingchoncharoen2013normal}.} 

\RP{All displacement nEPE and strain MAE comparisons are performed using two-sided Wilcoxon signed-rank tests~\cite{wilcoxon1992individual}. To account for the non-independence of slices from the same subject, we first average each metric across slices within each subject and then conduct paired comparisons using these subject-level values. We report the paired difference (GenStrain $-$ competing baseline) as the Hodges--Lehmann estimate~\cite{hodges2011estimates}, with its $95\%$ confidence interval derived from the Wilcoxon test. We also report the matched-pairs rank-biserial effect size~\cite{kerby2014simple} and the win rate, defined as the proportion of subjects for which GenStrain achieves lower error. To account for multiple comparisons, $p$-values are adjusted using the Holm correction~\cite{holm1979simple} with $p<0.05$ considered statistically significant.} 


We finally assess the agreement between strain estimates derived from the predicted motion and those measured by DENSE. We compute the intra-class correlation coefficient (ICC(2,1)) by using two-way random-effects model based on a single measurement for absolute agreement~\cite{shrout1979intraclass}. Pearson's CC and the coefficient of variation (CV) are reported to measure linear association~\cite{benesty2009pearson} and quantify relative variability~\cite{reed2002use}, respectively. Bland–Altman plots are used to assess systematic bias and limits of agreement (LoA) across test slices at the apical, mid-ventricular, and basal levels.

\subsection{Experimental Results}
\label{sec:results}
\begin{figure}[htbp]
\centering
\includegraphics[width=\linewidth]{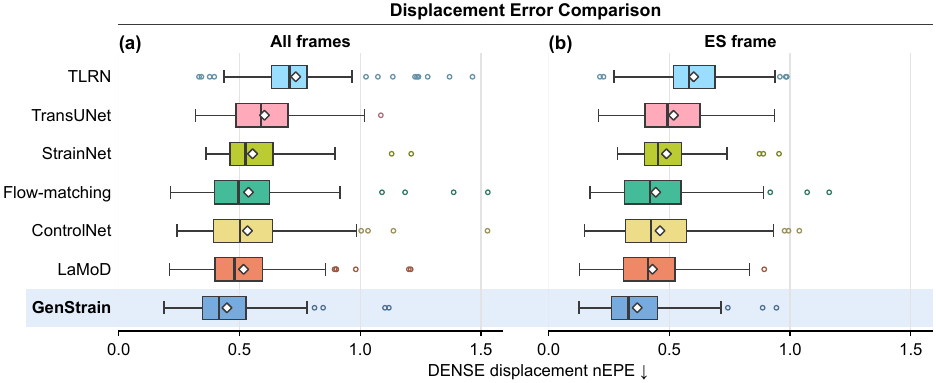}
\caption{\RP{Quantitative displacement error comparison on the DENSE test set. (a) All-frame and (b) end-systolic (ES) displacement nEPE. Boxes show the interquartile range, with the median (line) and mean (diamond).}}
\label{fig:nepe_media}
\end{figure}

 Fig.~\ref{fig:nepe_media} presents the nEPE across all frames (a) and at ES (b). It show that our method, GenStrain, consistently achieves the lowest displacement error among all baselines with statistically significant improvements. \RP{Detailed pairwise differences and Holm-corrected $p$-values are reported in (Tab.~\ref{tab:supp_displacement_significance})}.

\begin{table}[htbp]
\RPblue
\centering
\scriptsize
\setlength{\tabcolsep}{3pt}
\caption{Strain MAE (\%) on paired Cine--DENSE test data, across all frames and at end-systole (ES). DeepStrain is evaluated from cine input only, so the DENSE-input cells read ``--''. $^{*}$~significantly worse than GenStrain, $^{\dagger}$~significantly better, unmarked not significant. Per-comparison statistics: Tables~\ref{tab:supp_strain_dense}, \ref{tab:supp_strain_cine}.}
\label{tab:strain_error_metrics_media}
\begin{tabulary*}{\linewidth}{@{}LLCCCC@{}}
\toprule
 & & \multicolumn{2}{c}{\textbf{Whole-slice MAE} (\%) $\downarrow$} &
\multicolumn{2}{c}{\textbf{Segmental MAE} (\%) $\downarrow$} \\
\cmidrule(lr){3-4}\cmidrule(lr){5-6}
\textbf{Modality} & \textbf{Method} & \textbf{All-frame} & \textbf{ES} & \textbf{All-frame} & \textbf{ES} \\[1pt]
\midrule
\multirow{8}{*}{\makecell[l]{DENSE}} & DeepStrain & -- & -- & -- & -- \\
 & TLRN & 5.72 {\tiny$\pm$ 3.41}$^{*}$ & 6.55 {\tiny$\pm$ 4.10}$^{*}$ & 6.56 {\tiny$\pm$ 3.28}$^{*}$ & 8.06 {\tiny$\pm$ 3.83}$^{*}$ \\
 & TransUNet & 3.81 {\tiny$\pm$ 2.09}$^{*}$ & 5.51 {\tiny$\pm$ 3.78}$^{*}$ & 4.56 {\tiny$\pm$ 1.76}$^{*}$ & 6.88 {\tiny$\pm$ 3.48}$^{*}$ \\
 & StrainNet & 2.92 {\tiny$\pm$ 1.67}$^{*}$ & 4.56 {\tiny$\pm$ 2.84}$^{*}$ & 3.68 {\tiny$\pm$ 1.36}$^{*}$ & 5.51 {\tiny$\pm$ 2.35}$^{*}$ \\
 & Flow-matching & 2.73 {\tiny$\pm$ 2.25}$^{*}$ & 3.69 {\tiny$\pm$ 3.34}$^{*}$ & 3.69 {\tiny$\pm$ 2.04}$^{*}$ & 4.95 {\tiny$\pm$ 3.00}$^{*}$ \\
 & ControlNet & 2.71 {\tiny$\pm$ 1.83}$^{*}$ & 4.11 {\tiny$\pm$ 3.37}$^{*}$ & 3.49 {\tiny$\pm$ 1.59}$^{*}$ & 5.37 {\tiny$\pm$ 3.05}$^{*}$ \\
 & LaMoD & 2.59 {\tiny$\pm$ 1.61}$^{*}$ & 3.75 {\tiny$\pm$ 2.95}$^{*}$ & 3.13 {\tiny$\pm$ 1.46}$^{*}$ & 4.34 {\tiny$\pm$ 2.79}$^{*}$ \\
 & \cellcolor{lightgreen}GenStrain & \cellcolor{lightgreen}\textbf{2.14 {\tiny$\pm$ 1.56}} & \cellcolor{lightgreen}\textbf{2.81 {\tiny$\pm$ 2.43}} & \cellcolor{lightgreen}\textbf{2.73 {\tiny$\pm$ 1.46}} & \cellcolor{lightgreen}\textbf{3.68 {\tiny$\pm$ 2.26}} \\
\midrule
\multirow{8}{*}{\makecell[l]{Cine}} & DeepStrain & 4.94 {\tiny$\pm$ 2.09}$^{*}$ & 5.57 {\tiny$\pm$ 3.77}$^{*}$ & 5.58 {\tiny$\pm$ 1.82}$^{*}$ & 6.90 {\tiny$\pm$ 3.82}$^{*}$ \\
 & TLRN & 5.22 {\tiny$\pm$ 2.03}$^{*}$ & 5.98 {\tiny$\pm$ 3.26}$^{*}$ & 5.87 {\tiny$\pm$ 1.80}$^{*}$ & 7.70 {\tiny$\pm$ 3.51}$^{*}$ \\
 & TransUNet & 5.37 {\tiny$\pm$ 2.25}$^{*}$ & 7.84 {\tiny$\pm$ 3.95}$^{*}$ & 5.88 {\tiny$\pm$ 1.99}$^{*}$ & 9.21 {\tiny$\pm$ 3.70}$^{*}$ \\
 & StrainNet & 3.87 {\tiny$\pm$ 1.77}$^{*}$ & 5.55 {\tiny$\pm$ 3.06}$^{*}$ & 4.57 {\tiny$\pm$ 1.41}$^{*}$ & 7.08 {\tiny$\pm$ 2.96}$^{*}$ \\
 & Flow-matching & 3.96 {\tiny$\pm$ 2.44}$^{*}$ & 5.26 {\tiny$\pm$ 4.67}$^{*}$ & 5.22 {\tiny$\pm$ 2.19}$^{*}$ & 6.67 {\tiny$\pm$ 3.82}$^{*}$ \\
 & ControlNet & 4.42 {\tiny$\pm$ 2.30}$^{*}$ & 5.87 {\tiny$\pm$ 3.78}$^{*}$ & 4.87 {\tiny$\pm$ 2.07}$^{*}$ & 7.36 {\tiny$\pm$ 3.88}$^{*}$ \\
 & LaMoD & 3.69 {\tiny$\pm$ 1.72}$^{*}$ & 5.03 {\tiny$\pm$ 3.17}$^{*}$ & 4.22 {\tiny$\pm$ 1.46}$^{*}$ & 6.49 {\tiny$\pm$ 3.03}$^{*}$ \\
 & \cellcolor{lightgreen}GenStrain & \cellcolor{lightgreen}\textbf{2.84 {\tiny$\pm$ 1.57}} & \cellcolor{lightgreen}\textbf{2.90 {\tiny$\pm$ 2.39}} & \cellcolor{lightgreen}\textbf{3.53 {\tiny$\pm$ 1.43}} & \cellcolor{lightgreen}\textbf{4.76 {\tiny$\pm$ 2.86}} \\
\bottomrule
\end{tabulary*}
\end{table}

Tab.~\ref{tab:strain_error_metrics_media} reports strain MAE on both DENSE and cine CMR modalities. Overall, GenStrain consistently achieves the best strain estimation accuracy, with particularly substantial improvements on cine data and lower variability than competing methods. \RP{Its improvement over all baselines is statistically significant(Details are included in Appendix .~\ref{tab:supp_strain_dense},~\ref{tab:supp_strain_cine})}. 

\begin{figure}[htbp]
    \centering
    \begin{subfigure}[b]{0.9\textwidth}
        \centering
        \includegraphics[width=\textwidth]{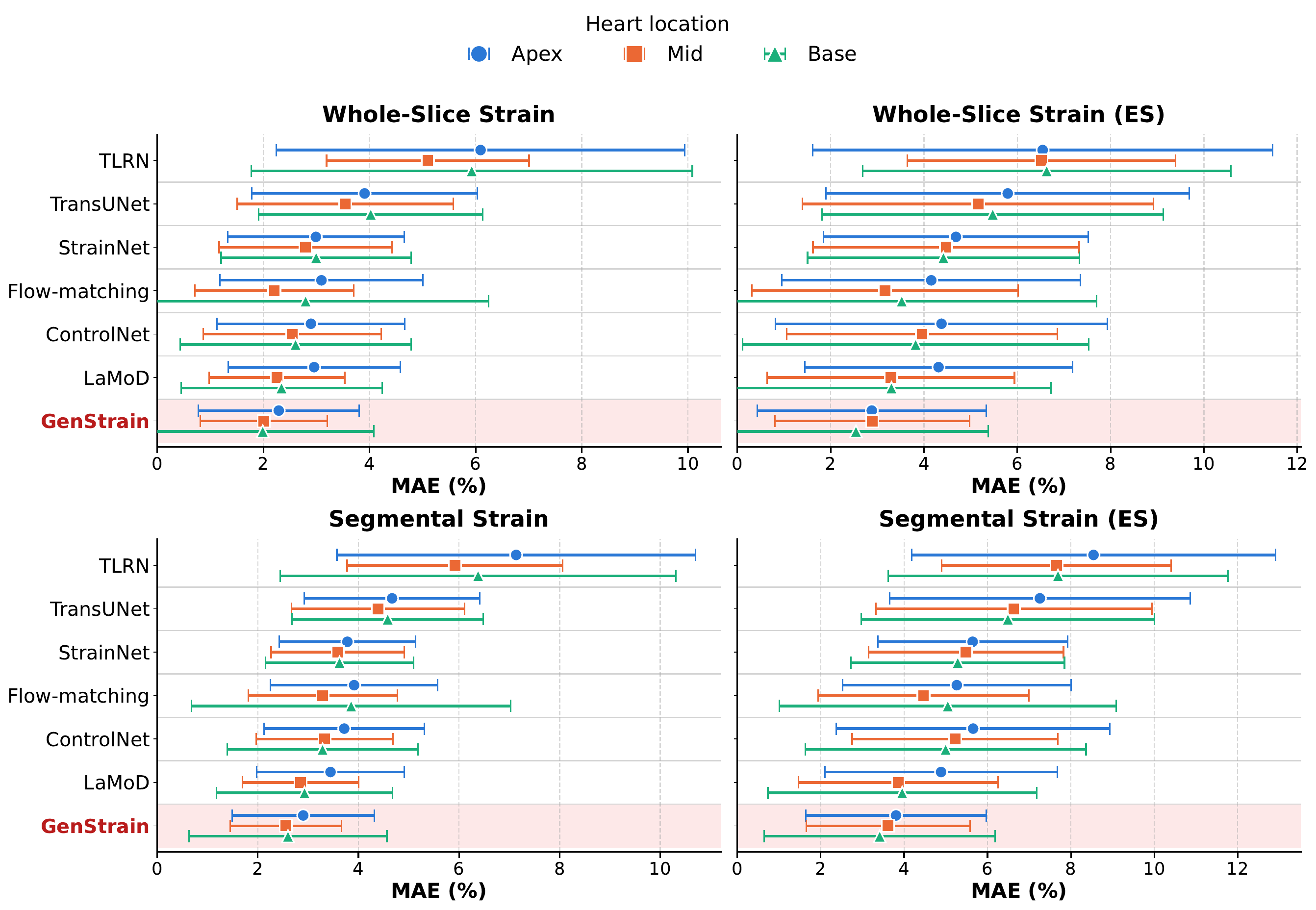}
        \caption{DENSE Test dataset}
        \label{fig:dense_strain_mae_barplots_comparison}
    \end{subfigure}
 
    \begin{subfigure}[b]{0.9\textwidth}
        \centering
        \includegraphics[width=\textwidth]{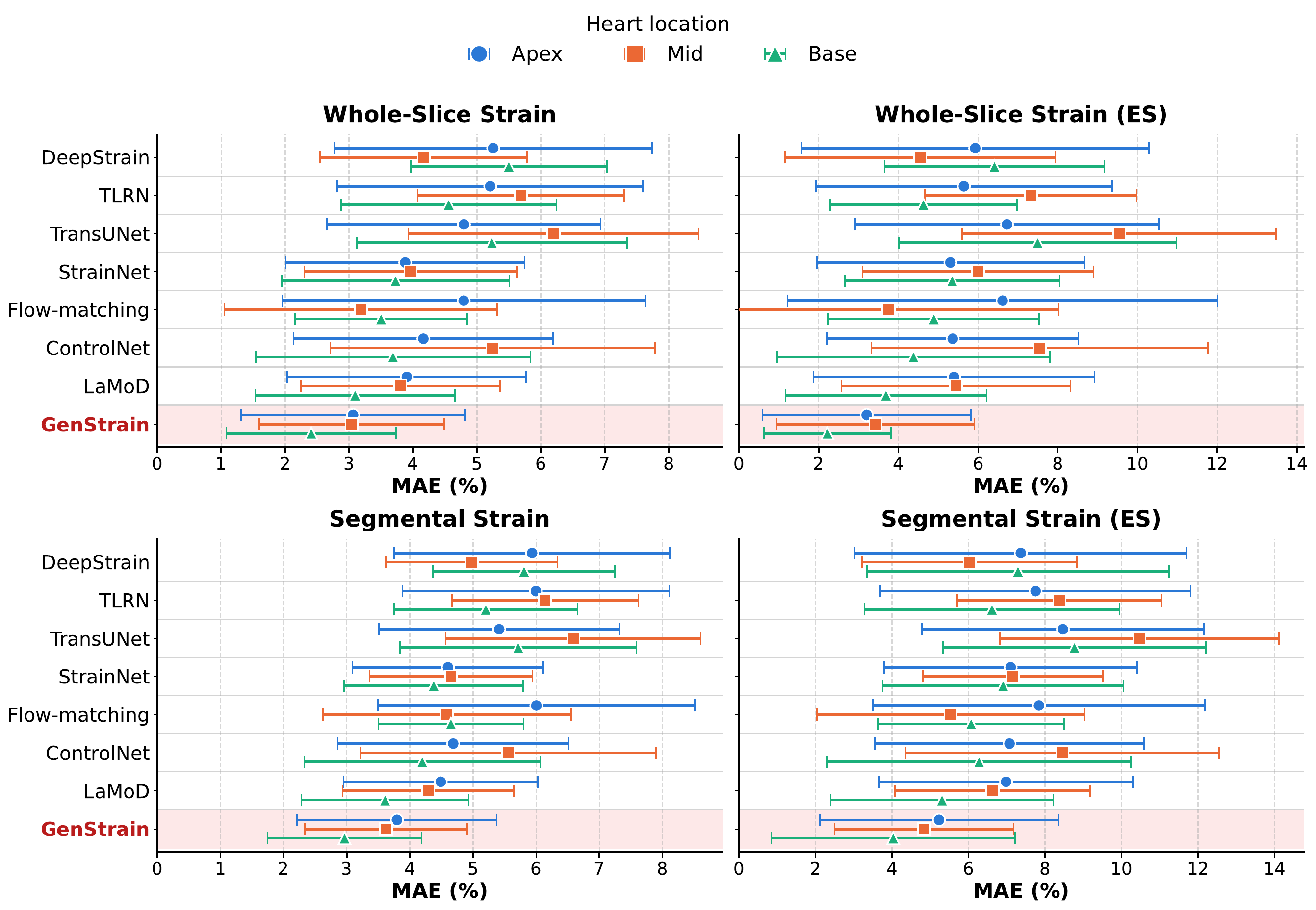}
        \caption{Cine Test dataset}
        \label{fig:cine_strain_mae_barplots_comparison}
    \end{subfigure}
    \caption{\RP{Comparison of Strain MAE (\%) across apical, mid-ventricular, and basal slices for all methods.}}
    \label{fig:strain_mae_barplots_comparison}
\end{figure}

Fig.~\ref{fig:strain_mae_barplots_comparison} presents comparisons of strain error across different cardiac regions (apex, mid, and base) for all evaluated models. \RP{It shows that our GenStrain achieves the lowest MAE across all slice locations in both DENSE and cine CMR data under full-frame and ES settings.} \RP{This suggests that GenStrain performs consistently across cardiac anatomical regions, which is important for reliable strain measurement independent of slice location.}

\begin{table}[!b]
\RPblue
\centering
\scriptsize
\setlength{\tabcolsep}{3pt}
\caption{Agreement between predicted and ground-truth strain, with 95\% confidence intervals. Higher is better for ICC and Pearson CC, and lower is better for CV. Note that DeepStrain results are reported only for cine inputs as corresponding DENSE magnitude images are not available.}
\label{tab:agreement_media}
\begin{tabulary*}{\linewidth}{@{}LLCcCC@{}}
\toprule
\textbf{Cohort} & \textbf{Method} & \textbf{ICC} $\uparrow$ & \textbf{Pearson CC} $\uparrow$ & \textbf{CV} (\%) $\downarrow$ & \makecell{\textbf{Bias}\\$\pm$ \textbf{95\% LOA}} \\[1pt]
\midrule
\multirow{8}{*}{\makecell[l]{DENSE\\Whole-slice}} & DeepStrain & -- & -- & -- & -- \\
 & TLRN & 0.39 {\tiny [0.28, 0.57]} & 0.52 {\tiny [0.41, 0.82]} & 108.26 {\tiny [48.53, 163.74]} & 5.63 {\tiny $\pm$ 15.47} \\
 & TransUNet & 0.65 {\tiny [0.58, 0.72]} & 0.79 {\tiny [0.74, 0.84]} & 44.73 {\tiny [39.33, 51.07]} & 3.32 {\tiny $\pm$ 7.40} \\
 & StrainNet & 0.77 {\tiny [0.72, 0.82]} & \textbf{0.89 {\tiny [0.85, 0.92]}} & 33.84 {\tiny [28.99, 39.08]} & 2.39 {\tiny $\pm$ 5.91} \\
 & Flow-matching & 0.80 {\tiny [0.70, 0.86]} & 0.81 {\tiny [0.72, 0.87]} & 39.96 {\tiny [31.90, 50.75]} & 1.07 {\tiny $\pm$ 7.49} \\
 & ControlNet & 0.80 {\tiny [0.73, 0.85]} & 0.85 {\tiny [0.80, 0.89]} & 35.81 {\tiny [30.23, 41.97]} & 1.87 {\tiny $\pm$ 6.43} \\
 & LaMoD & 0.83 {\tiny [0.77, 0.87]} & 0.84 {\tiny [0.78, 0.88]} & 35.26 {\tiny [29.93, 41.54]} & 0.55 {\tiny $\pm$ 6.79} \\
 & \cellcolor{lightgreen}GenStrain & \cellcolor{lightgreen}\textbf{0.88 {\tiny [0.83, 0.92]}} & \cellcolor{lightgreen}0.88 {\tiny [0.84, 0.92]} & \cellcolor{lightgreen}\textbf{29.89 {\tiny [24.36, 36.46]}} & \cellcolor{lightgreen}0.58 {\tiny $\pm$ 5.75} \\
\midrule
\multirow{8}{*}{\makecell[l]{DENSE\\Segmental}} & DeepStrain & -- & -- & -- & -- \\
 & TLRN & 0.34 {\tiny [0.25, 0.48]} & 0.44 {\tiny [0.35, 0.64]} & 121.75 {\tiny [75.80, 168.93]} & 5.58 {\tiny $\pm$ 17.38} \\
 & TransUNet & 0.58 {\tiny [0.52, 0.64]} & 0.69 {\tiny [0.63, 0.73]} & 58.79 {\tiny [52.48, 66.12]} & 3.35 {\tiny $\pm$ 9.67} \\
 & StrainNet & 0.68 {\tiny [0.63, 0.73]} & 0.77 {\tiny [0.73, 0.81]} & 47.74 {\tiny [42.77, 53.33]} & 2.43 {\tiny $\pm$ 8.29} \\
 & Flow-matching & 0.70 {\tiny [0.57, 0.78]} & 0.71 {\tiny [0.59, 0.79]} & 53.91 {\tiny [43.78, 69.60]} & 1.16 {\tiny $\pm$ 10.03} \\
 & ControlNet & 0.72 {\tiny [0.66, 0.77]} & 0.76 {\tiny [0.71, 0.80]} & 48.15 {\tiny [42.45, 54.77]} & 1.92 {\tiny $\pm$ 8.60} \\
 & LaMoD & 0.79 {\tiny [0.72, 0.83]} & 0.79 {\tiny [0.73, 0.84]} & 42.45 {\tiny [37.08, 49.67]} & 0.51 {\tiny $\pm$ 8.17} \\
 & \cellcolor{lightgreen}GenStrain & \cellcolor{lightgreen}\textbf{0.83 {\tiny [0.76, 0.87]}} & \cellcolor{lightgreen}\textbf{0.83 {\tiny [0.77, 0.87]}} & \cellcolor{lightgreen}\textbf{38.48 {\tiny [32.41, 46.38]}} & \cellcolor{lightgreen}0.62 {\tiny $\pm$ 7.36} \\
\midrule
\multirow{8}{*}{\makecell[l]{Cine\\Whole-slice}} & DeepStrain & 0.69 {\tiny [0.64, 0.74]} & 0.77 {\tiny [0.70, 0.83]} & 60.37 {\tiny [53.67, 67.76]} & -0.75 {\tiny $\pm$ 12.18} \\
 & TLRN & 0.57 {\tiny [0.50, 0.63]} & 0.75 {\tiny [0.66, 0.82]} & 55.66 {\tiny [46.18, 66.29]} & 4.65 {\tiny $\pm$ 8.29} \\
 & TransUNet & 0.44 {\tiny [0.36, 0.51]} & 0.66 {\tiny [0.56, 0.75]} & 61.34 {\tiny [53.64, 70.47]} & 4.89 {\tiny $\pm$ 8.99} \\
 & StrainNet & 0.62 {\tiny [0.56, 0.68]} & \textbf{0.81 {\tiny [0.76, 0.86]}} & 44.48 {\tiny [39.58, 49.74]} & 3.45 {\tiny $\pm$ 7.15} \\
 & Flow-matching & 0.73 {\tiny [0.65, 0.80]} & 0.79 {\tiny [0.71, 0.85]} & 51.12 {\tiny [42.73, 59.52]} & -1.30 {\tiny $\pm$ 10.59} \\
 & ControlNet & 0.56 {\tiny [0.48, 0.64]} & 0.71 {\tiny [0.63, 0.79]} & 53.97 {\tiny [46.27, 62.33]} & 3.84 {\tiny $\pm$ 8.46} \\
 & LaMoD & 0.64 {\tiny [0.58, 0.70]} & 0.81 {\tiny [0.75, 0.86]} & 43.55 {\tiny [38.72, 49.37]} & 3.18 {\tiny $\pm$ 7.12} \\
 & \cellcolor{lightgreen}GenStrain & \cellcolor{lightgreen}\textbf{0.78 {\tiny [0.72, 0.84]}} & \cellcolor{lightgreen}0.81 {\tiny [0.75, 0.86]} & \cellcolor{lightgreen}\textbf{40.36 {\tiny [34.41, 46.74]}} & \cellcolor{lightgreen}1.47 {\tiny $\pm$ 7.27} \\
\midrule
\multirow{8}{*}{\makecell[l]{Cine\\Segmental}} & DeepStrain & 0.63 {\tiny [0.57, 0.68]} & 0.68 {\tiny [0.61, 0.74]} & 68.62 {\tiny [61.98, 76.19]} & -0.83 {\tiny $\pm$ 13.88} \\
 & TLRN & 0.50 {\tiny [0.43, 0.55]} & 0.62 {\tiny [0.55, 0.68]} & 75.56 {\tiny [67.20, 85.25]} & 4.71 {\tiny $\pm$ 11.18} \\
 & TransUNet & 0.41 {\tiny [0.33, 0.47]} & 0.56 {\tiny [0.48, 0.63]} & 77.21 {\tiny [69.39, 86.44]} & 5.05 {\tiny $\pm$ 11.17} \\
 & StrainNet & 0.52 {\tiny [0.47, 0.58]} & 0.67 {\tiny [0.61, 0.71]} & 60.30 {\tiny [55.18, 65.73]} & 3.47 {\tiny $\pm$ 9.66} \\
 & Flow-matching & 0.64 {\tiny [0.57, 0.70]} & 0.69 {\tiny [0.63, 0.75]} & 65.77 {\tiny [58.28, 74.07]} & -1.22 {\tiny $\pm$ 13.55} \\
 & ControlNet & 0.53 {\tiny [0.44, 0.61]} & 0.65 {\tiny [0.57, 0.73]} & 64.49 {\tiny [56.09, 73.80]} & 3.94 {\tiny $\pm$ 10.03} \\
 & LaMoD & 0.60 {\tiny [0.55, 0.65]} & 0.73 {\tiny [0.68, 0.77]} & 54.12 {\tiny [49.29, 59.82]} & 3.16 {\tiny $\pm$ 8.83} \\
 & \cellcolor{lightgreen}GenStrain & \cellcolor{lightgreen}\textbf{0.73 {\tiny [0.67, 0.78]}} & \cellcolor{lightgreen}\textbf{0.75 {\tiny [0.70, 0.80]}} & \cellcolor{lightgreen}\textbf{49.11 {\tiny [43.19, 55.66]}} & \cellcolor{lightgreen}1.49 {\tiny $\pm$ 8.82} \\
\bottomrule
\end{tabulary*}
\end{table}

\begin{figure}[htbp]
\centering
\includegraphics[width=\textwidth]{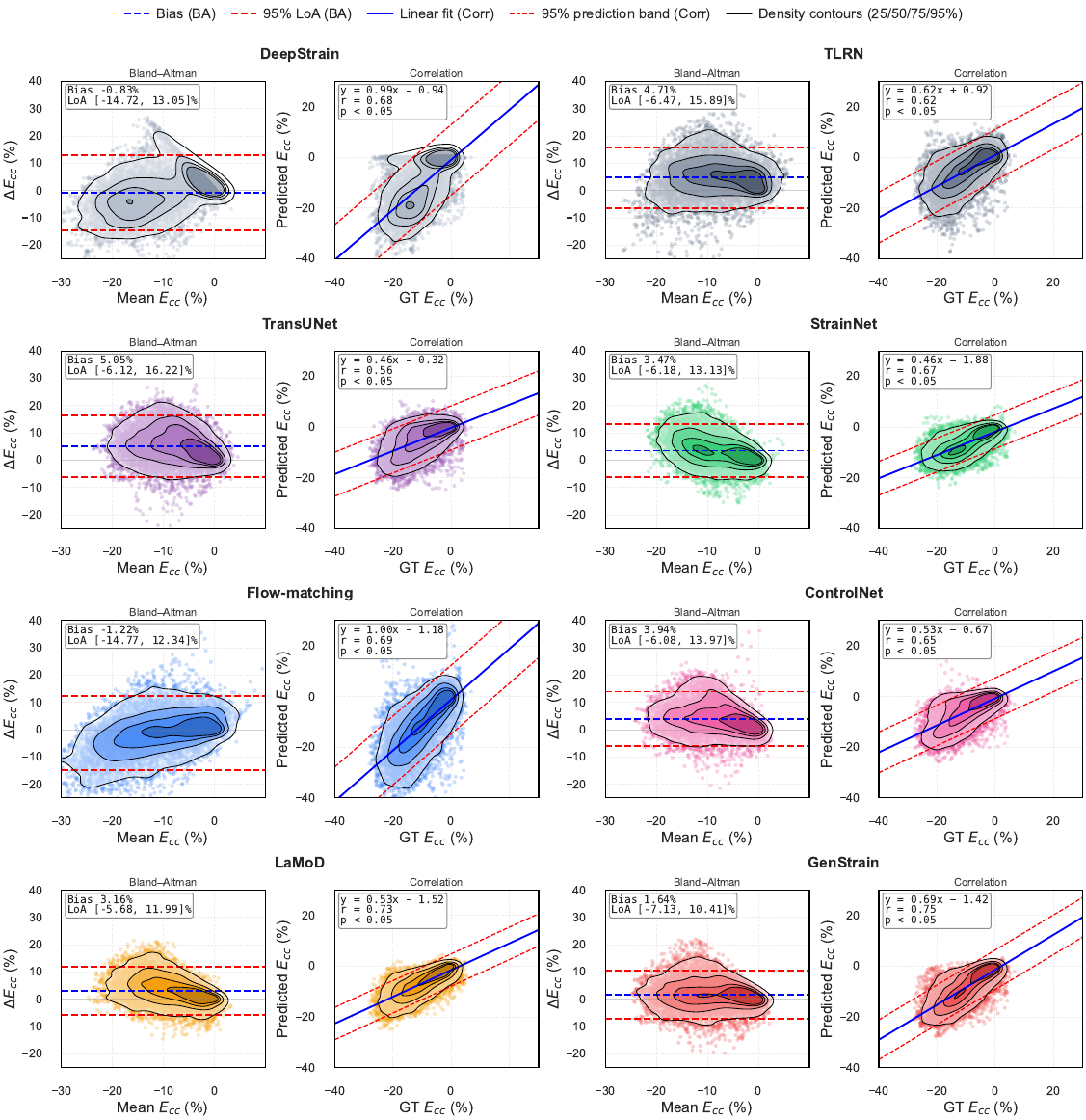}
\caption{\RP{Visualization of Bland--Altman (left) and correlation (right) plots for cine segmental strain vs. DENSE. The density contours enclose 25/50/75/95\% of the data points.}}
\label{fig:ba-corr-segmental-cine}
\end{figure}

Tab.~\ref{tab:agreement_media} assesses the agreement between model-predicted strain and DENSE-strain on paired cine-DENSE CMR test data. Overall, GenStrain agrees the best with DENSE-derived strain, with the highest ICC and lowest CV in all four strain metrics. \RP{The baseline StrainNet~\cite{wang2023strainnet} and LaMoD~\cite{xing2024lamod} achieve Pearson correlations comparable to GenStrain on cine input ($r=0.81$) but exhibit substantially lower ICCs due to systematic biases of 3--5\%, a similar limitation observed in other methods, including TLRN~\cite{wu2024tlrn}, TransUNet~\cite{chen2021transunet}, and ControlNet~\cite{zhang2023adding}. In contrast, DeepStrain~\cite{morales2021deepstrain} and Flow Matching~\cite{lipman2023flow} exhibit minimal systematic bias but substantially wider limits of agreement, indicating reduced reliability at the individual-measurement level. In Fig.~\ref{fig:ba-corr-segmental-cine}, we visualize Bland--Altman and correlation plots on the segmental strain of the cine dataset. GenStrain achieves the strongest agreement with DENSE ground truth, with the highest correlation ($r = 0.75$), the narrowest LoA ($[-7.13, 10.41]\%$), and a low bias ($1.64\%$), indicating that its segmental errors are both small and consistent.}

\begin{figure}[htbp]
\centering
\includegraphics[width=\textwidth]{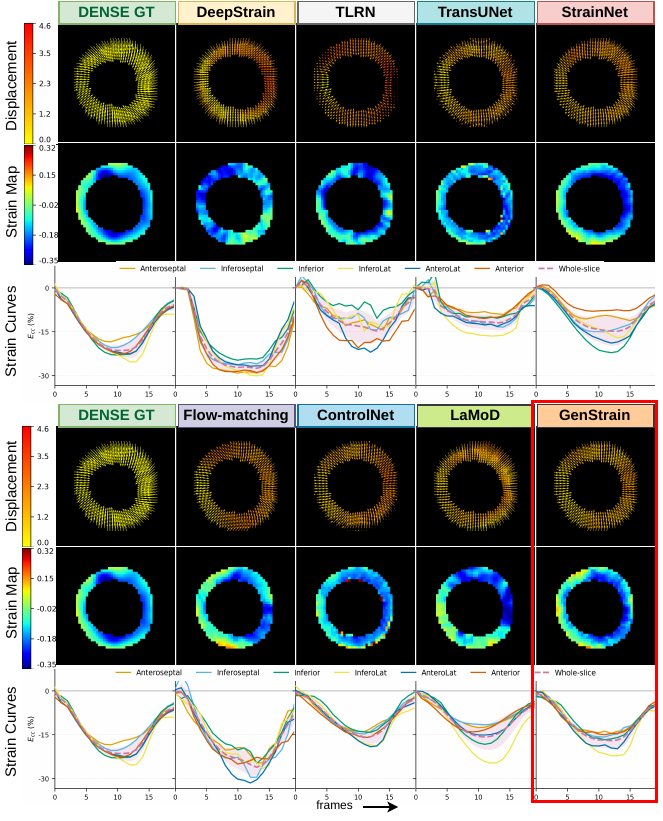}
    \caption{\RP{Healthy Subject example: ES displacement (top), circumferential strain map (middle), and segmental/whole-slice strain curve (bottom). GenStrain better aligns with DENSE GT, with yellow arrows indicating better alignment.}}
    \label{fig:healthy_example}
\end{figure}

\begin{figure}[htbp]
\centering
\includegraphics[width=\textwidth]{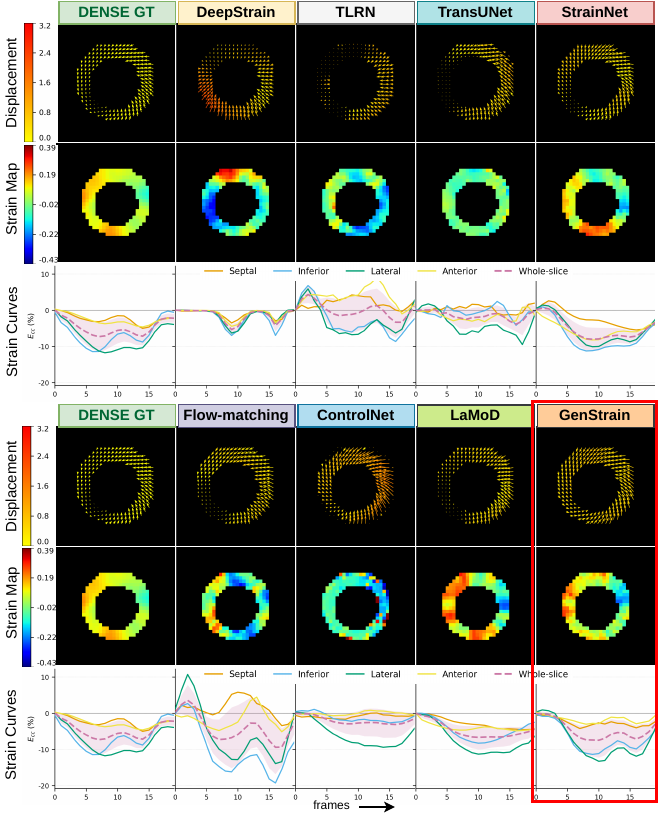}
    \caption{\RP{Heart disease (Acute MI Patient) example: ES displacement (top), circumferential strain map (middle), and segmental/whole-slice strain curve (bottom). GenStrain better aligns with DENSE GT, with yellow arrows indicating better alignment.}}
    \label{fig:disease_example}
\end{figure}

Fig.~\ref{fig:healthy_example} and Fig.~\ref{fig:disease_example} visualize two examples (healthy subject vs.\ acute myocardial infarction (MI) patient) comparing ES displacement fields, $E_{cc}$ strain maps, and segmental strain curves across all methods. Our method, GenStrain, achieves the closest agreement with the DENSE across all segments throughout the cardiac cycle. Similarly, the ES displacement, color-coded from yellow (high similarity) to red (low similarity) relative to DENSE, appears to show the highest similarity.

\paragraph*{\bf Ablation Study} To investigate the contribution of each condition to our multi-conditional diffusion model, we performed an ablation study on the paired Cine--DENSE test set by selectively removing the contour or motion condition illustrated in Tab.~\ref{tab:ablation_strain_error_metrics}. \RP{It shows that both conditioning can provide complementary information that benefits the strain estimation across all error metrics. However, it is worth noting that severe motion errors from the spatiotemporal network can affect GenStrain’s performance. However, GenStrain can flexibly chose to reduce its reliance on motion conditioning, or rely solely on contour conditioning if necessary.}

\begin{table}[htbp]
\RPblue
\centering
\scriptsize
\setlength{\tabcolsep}{3pt}
\caption{Ablation on different combination of conditioning.}
\label{tab:ablation_strain_error_metrics}
\begin{tabulary*}{\linewidth}{@{}LLCCCCCC@{}}
\toprule
 &  & \multicolumn{2}{c}{\textbf{Condition}} &
\multicolumn{2}{c}{\makecell{\textbf{Whole-slice}\\\textbf{MAE} (\%) $\downarrow$}} &
\multicolumn{2}{c}{\makecell{\textbf{Segmental}\\\textbf{MAE} (\%) $\downarrow$}} \\
\cmidrule(lr){3-4}\cmidrule(lr){5-6}\cmidrule(lr){7-8}
\textbf{Modality} & \textbf{Variant} & \textbf{Contour} & \textbf{Motion} & \textbf{All-frame} & \textbf{ES} & \textbf{All-frame} & \textbf{ES} \\[1pt]
\midrule
\multirow{3}{*}{\makecell[l]{\textbf{DENSE}}}
 & Contour only & \cmark & \xmark & 2.18 {\tiny$\pm$ 1.59} & 2.91 {\tiny$\pm$ 2.59} & 2.77 {\tiny$\pm$ 1.48} & \textbf{3.68 {\tiny$\pm$ 2.44}} \\
 & Motion only & \xmark & \cmark & 2.25 {\tiny$\pm$ 1.56} & 3.05 {\tiny$\pm$ 2.70} & 2.92 {\tiny$\pm$ 1.40} & 3.97 {\tiny$\pm$ 2.54} \\
 & \textbf{GenStrain} & \cmark & \cmark & \textbf{2.14 {\tiny$\pm$ 1.56}} & \textbf{2.81 {\tiny$\pm$ 2.43}} & \textbf{2.73 {\tiny$\pm$ 1.46}} & 3.68 {\tiny$\pm$ 2.26} \\
\midrule
\multirow{3}{*}{\makecell[l]{\textbf{Cine}}}
 & Contour only & \cmark & \xmark & \textbf{2.84 {\tiny$\pm$ 1.54}} & 3.17 {\tiny$\pm$ 2.39} & \textbf{3.52 {\tiny$\pm$ 1.40}} & 4.94 {\tiny$\pm$ 2.74} \\
 & Motion only & \xmark & \cmark & 3.09 {\tiny$\pm$ 1.63} & 3.47 {\tiny$\pm$ 2.59} & 3.79 {\tiny$\pm$ 1.50} & 5.17 {\tiny$\pm$ 2.90} \\
 & \textbf{GenStrain} & \cmark & \cmark & 2.84 {\tiny$\pm$ 1.57} & \textbf{2.90 {\tiny$\pm$ 2.39}} & 3.53 {\tiny$\pm$ 1.43} & \textbf{4.76 {\tiny$\pm$ 2.86}} \\
\bottomrule
\end{tabulary*}
\end{table}

 
\subsection{Assess Translational Potential} 
\RP{We explore the potential for clinical translation of GenStrain by
benchmarking its performance against a widely used commercial FT software (NeoSoft's suiteHEART) on an independent set of $91$ cine CMR slices from $34$ subjects. The FT analysis was performed by a clinical expert with careful adjustment on the segmented LV contours from the software. Both methods are evaluated against DENSE-derived strain from paired acquisitions at the matched location. }
 
\RP{Tab.~\ref{tab:ft_vs_genstrain_cine} reports the quantitative comparison of whole-slice and segmental strain MAE between FT and GenStrain, using DENSE as the reference. While GenStrain achieves comparable performance to FT in whole-slice strain estimation, it demonstrates substantially improved accuracy in segmental strain, particularly in diseased cases.}
\begin{table}[htbp]
\RPblue
\centering
\scriptsize
\setlength{\tabcolsep}{3pt}
\caption{Comparison of strain MAE ($\%$) for GenStrain and FT against DENSE on the cine test set, evaluated across all frames and ES. The last row reports Holm-adjusted $p$-values for paired comparison; the $^{*}$~indicates GenStrain achieves significantly better results in segmental strain.}
\label{tab:ft_vs_genstrain_cine}
\begin{tabulary*}{\linewidth}{@{}LCCCC@{}}
\toprule
 & \multicolumn{2}{c}{\textbf{Whole-slice MAE} (\%) $\downarrow$} &
\multicolumn{2}{c}{\textbf{Segmental MAE} (\%) $\downarrow$} \\
\cmidrule(lr){2-3}\cmidrule(lr){4-5}
\textbf{Method} & \textbf{All-frame} & \textbf{ES} & \textbf{All-frame} & \textbf{ES} \\[1pt]
\midrule
Feature Tracking & \mbox{2.84 {\tiny$\pm$ 1.46}} & \mbox{2.86 {\tiny$\pm$ 2.54}} & \mbox{4.69 {\tiny$\pm$ 1.34}}$^{*}$ & \mbox{5.97 {\tiny$\pm$ 2.59}}$^{*}$ \\
\textbf{GenStrain} & \textbf{\mbox{2.78 {\tiny$\pm$ 1.49}}} & \textbf{\mbox{2.62 {\tiny$\pm$ 1.95}}} & \textbf{\mbox{3.46 {\tiny$\pm$ 1.39}}} & \textbf{\mbox{4.60 {\tiny$\pm$ 2.84}}} \\
\midrule
\textit{Holm $p$} & 1.000 & 1.000 & \textbf{\boldmath$9.4 \times 10^{-5}$} & \textbf{\boldmath$6.5 \times 10^{-3}$} \\
\bottomrule
\end{tabulary*}
\end{table}

\begin{figure}[!h]
\centering
\includegraphics[width=0.78\textwidth]{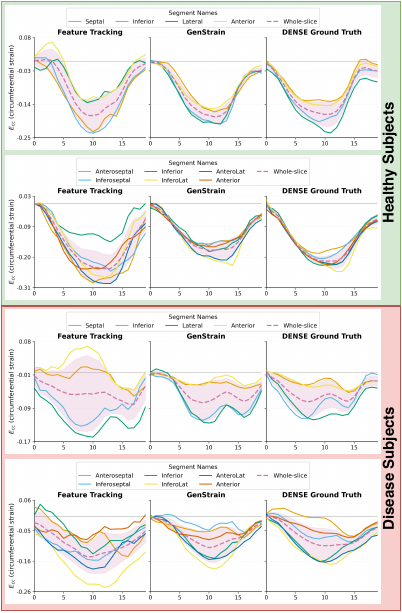}
    \caption{\RP{Qualitative comparison of circumferential strain curves across cardiac phases for healthy (top two rows) and disease (bottom two rows) subjects. GenStrain demonstrates closer agreement with DENSE ground truth for both whole-slice and segmental strain, whereas Feature Tracking exhibits larger deviations, particularly in segmental strain.}}
    \label{fig:ft_examples}
\end{figure}

\RP{Fig.~\ref{fig:ft_examples} presents representative strain curves from FT, GenStrain, and DENSE for both healthy subjects and patients with myocardial infarction.}

\RP{While we acknowledge this comparison was conducted on a relatively small subset of cine CMR images with available FT measurements, our results show a promising value of potential clinical translation of GenStrain. A comprehensive clinical validation using larger and more diverse cohorts will be an important direction for future work.}

\begin{table}[!t]
\RPblue
\centering
\setlength{\tabcolsep}{3pt}
\caption{Computational cost on the cine test set. Inference time is the average time per slice (a 20-frame sequence).}
\label{tab:computational_cost}
\resizebox{\linewidth}{!}{%
\begin{tabular}{@{}lccccccccc@{}}
\toprule
\textbf{Method}
  & \makecell{Feature\\Tracking}
  & DeepStrain
  & TLRN
  & TransUNet
  & StrainNet
  & \makecell{Flow-\\matching}
  & ControlNet
  & LaMoD
  & \textbf{GenStrain} \\
\midrule
\makecell[l]{\textbf{Inference}\\\textbf{time} (s) $\downarrow$}
  & 20.00 & 1.39 & 0.12 & 0.03 & 0.004 & 7.45 & 3.64 & 0.30 & 3.46 \\
\addlinespace
\makecell[l]{\textbf{Allocated GPU}\\\textbf{memory} (GB) $\downarrow$}
  & --   & 0.77 & 0.02 & 0.51 & 0.02  & 1.77 & 0.85  & 0.48 & 0.67 \\
\bottomrule
\end{tabular}%
}
\end{table}

\paragraph*{\bf Computational Cost}
\RP{We then evaluated GenStrain’s computational cost to assess its feasibility for clinical translation. Using $50$-step DDIM sampling~\cite{song2020denoising}, the inference of GenStrain takes $3.46$\,s per slice and requires $0.67$\,GB of GPU memory. As shown in Tab.~\ref{tab:computational_cost}, GenStrain achieves faster inference than most competing generative models and commercial FT ($20$s). Please note the reported FT runtime includes the average strain analysis time across basal, mid-ventricular, and apical slices. GenStrain’s fast inference and modest GPU memory requirements highlight its potential for practical clinical use.}

\begin{figure}[!h]
\centering
\includegraphics[width=\textwidth]{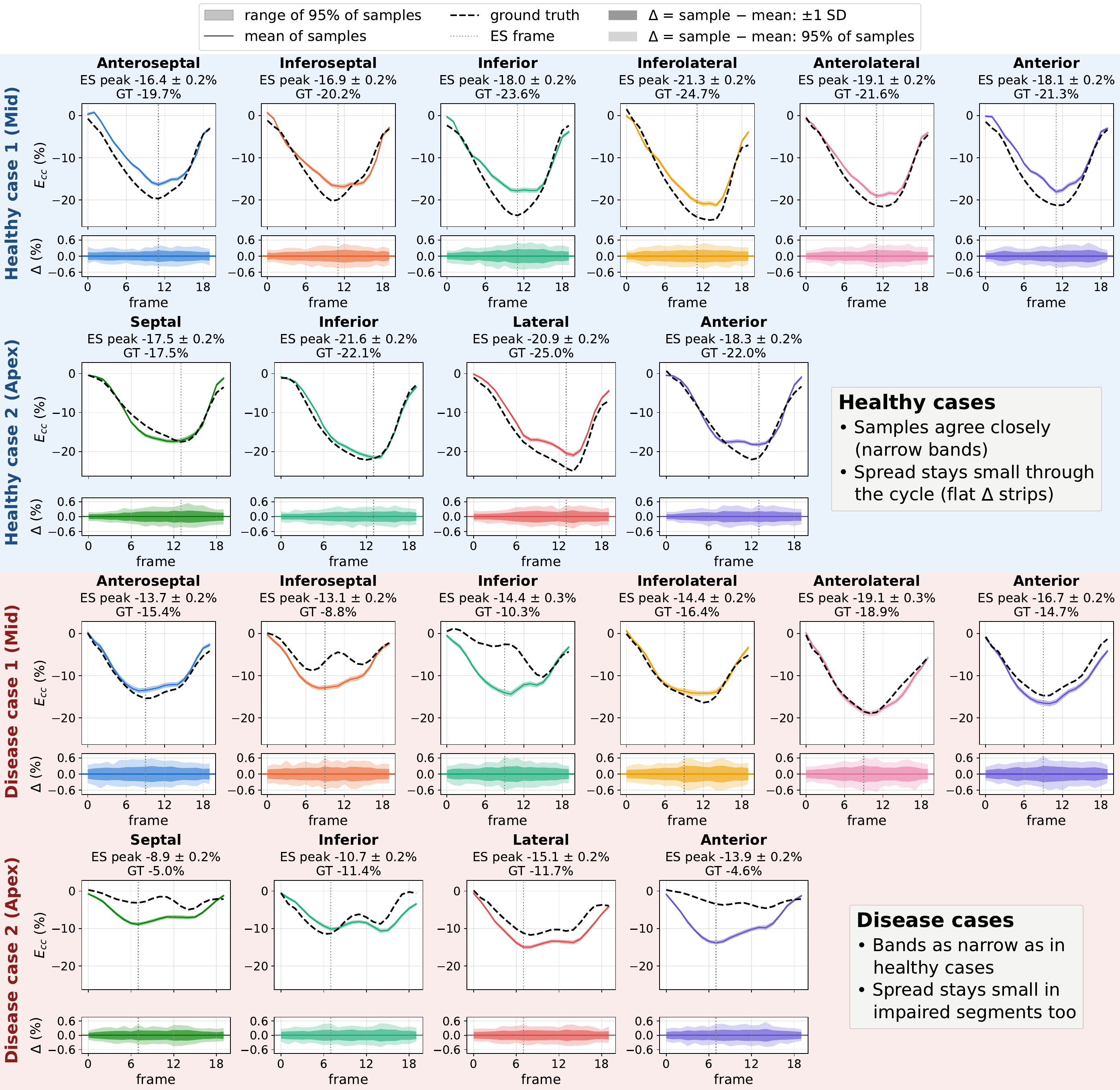}
    \caption{\RP{Top to bottom: Examples of segmental $E_{cc}$ uncertainty for two healthy and two acute-infarct test slices (mean $\pm$ SD). The strips below each panel show the distribution of $95\%$ samples around the mean.}}
    \label{fig:uncertainty}
\end{figure}

\RP{\paragraph*{\bf Strain Uncertainty} We further quantified the uncertainty in its strain estimates, as reliability is critical for clinical decision-making. Specifically, we generated $100$ motion samples per cine input and measured the variability of segmental $E_{cc}$ strain across samples. Fig.~\ref{fig:uncertainty} displays examples of segmental $E_{cc}$ uncertainty on both healthy and disease subjects. The relatively narrow uncertainty intervals indicate that GenStrain produces consistent strain estimates across samples under the same input conditioning. With a decent agreement with DENSE, these results indicate the potential reliability of GenStrain for downstream clinical applications.}

\providecommand{\dm}[1]{\textcolor{gray}{#1}}

\section{Conclusion \& Discussion}
\label{sec:discussion}

In this paper, we introduce GenStrain, a new multi-conditional diffusion framework for myocardial motion generation with improved accuracy in strain analysis of routinely acquired cine CMRs. By jointly leveraging temporal motion dynamics and CMR-derived anatomical features through a spatio-temporal cross-attention mechanism, our model effectively learns the complex distribution of high-fidelity motion patterns typically available only from advanced DENSE imaging modalities. Experimental results demonstrate that GenStrain achieves state-of-the-art performance in myocardial strain estimation from cine CMR. \RP{Beyond methodological contributions, we explore GenStrain's potential for translational value. By enabling accurate strain estimation directly from routine cine MRI without requiring dedicated acquisition protocols, our model has the potential to streamline clinical workflows, facilitate more efficient detection of cardiac dysfunction, and support personalized treatment strategies. Despite its effectiveness, GenStrain has several limitations that warrant further investigation and improvement as additional data become available in future works.}

\paragraph*{\bf Failure modes} \RP{The proposed motion-conditioning strategy may not fully eliminate severe registration errors. In challenging cases, such as patients with LBBB or substantial regional pathological changes, inaccurate motion priors may propagate into the generative process and lead to erroneous motion estimates. Furthermore, the current image conditioning relies on structural anatomical information derived from DENSE images, as their magnitude images provide limited contrast for capturing detailed tissue textures. This may constrain GenStrain’s ability to learn complex tangential motion patterns, particularly large rotations associated with twist or torsion~\cite{lei2025dense}. These factors highlight potential failure modes and motivate our future investigation into improving GenStrain’s robustness to inaccurate motion priors and complex pathological motion.}

\paragraph*{\bf Model generalizability} \RP{Improving model generalizability is an important direction for future development of GenStrain. Evaluating its performance on unseen or rare cardiac diseases with distinct motion patterns, as well as on substantially different patient populations (e.g., training on adults and testing on young children), is critical to assess its robustness to distribution shifts in DENSE-derived motion patterns. Such shifts may challenge the model’s ability to generate accurate motion estimates, particularly in underrepresented clinical populations. A more comprehensive study of generalizability across multi-clinical sites, imaging protocols, and scanner vendors is part of our ongoing work and is beyond the scope of the current study.}   

\paragraph*{\bf Enhancing clinical utility} \RP{Our validation of GenStrain has primarily focused on motion and strain estimation accuracy, representing a promising first step toward clinical translation. Further validation in diverse clinical settings is needed to establish its clinical utility and reliability. Rigorous validation of regional strain-map fidelity is important for establishing clinical utility. In particular, the predicted regional strain patterns would be best to be validated against independent evidence, such as regional dysfunction labels (i.e., late gadolinium enhancement scar extent), or clinical outcomes. However, assembling and curating the necessary datasets, including reliable regional reference labels and corresponding clinical information, is an ongoing effort that requires substantial time and resources. As the primary objective of the present study is to establish and evaluate the proposed methodological framework, this additional validation represents a critical next step toward clinical translation, which we plan to pursue as the required datasets become available.}

\paragraph*{\bf Acknowledgment} This work was supported by NIH R01HL181107, NIH 1R21EB032597, NSF CAREER Grant 2239977, and funding from the UVA Wallace H. Coulter Center for Translational Research. We gratefully acknowledge the data contribution from Daniel C. Auger, PhD; Pierre Croisille, MD, PhD; Magalie Viallon, PhD; Kenneth Mangion, MD, PhD; Colin Berry, MD, PhD; Christopher M. Haggerty, PhD; Lingyuan Jing, PhD; Brandon K. Fornwalt, MD, PhD; J. Jane Cao, MD, MPH; Joshua Cheng, BSc; Andrew D. Scott, PhD; Pedro F. Ferreira, PhD; John N. Oshinski, PhD; Daniel B. Ennis, PhD; Kenneth C. Bilchick, MD, MSc.

\newpage
\appendix

\section{Per-comparison Statistical Results} 
\label{app:stats}
\RP{This appendix provides the full pairwise statistical analysis of experimental comparisons summarized in Sec.~\ref{sec:experiments}. For each error metric, we report paired differences with $95\%$ confidence intervals, effect sizes, and Holm-corrected $p$-values, comparing our GenStrain with baseline methods or evaluating different conditioning configurations.} 

\RP{Tab.~\ref{tab:supp_displacement_significance} provides the statistical analysis of the displacement nEPE results (presented in main paper Fig.~\ref{fig:nepe_media}), comparing GenStrain with the baseline methods on DENSE inputs.}
\begin{table}[htbp]
\RPblue
\centering
\scriptsize
\setlength{\tabcolsep}{3pt}
\caption{Statistical results for DENSE nEPE corresponding to Fig.~\ref{fig:nepe_media} in the main paper. An $*$ indicates a baseline with significantly higher error than GenStrain after Holm correction across 12 comparisons.}
\label{tab:supp_displacement_significance}
\begin{tabulary*}{\linewidth}{@{}LLCcCC@{}}
\toprule
 & & & \multicolumn{2}{c}{\makecell{\textbf{Paired difference}\\\textbf{(GenStrain $-$ baseline)}}} & \\
\cmidrule(lr){4-5}
\textbf{Metric} & \textbf{Baseline} & \makecell{\textbf{Baseline}\\\textbf{(mean)}} & \makecell{\textbf{HL estimate}\\\textbf{[95\% CI]}} & \makecell{\textbf{Effect size}\\\textbf{(win rate)}} & \makecell{\textbf{Holm}\\$p$} \\[1pt]
\midrule
\multirow{6}{*}{\makecell[l]{All-frame\\nEPE\\[2pt]\textbf{GenStrain: 0.478}}} & TLRN & 0.767 & \mbox{-0.287 {\tiny [-0.332, -0.244]}} & 0.99 (96\%) & \textbf{\boldmath$2.1 \times 10^{-13}\,$}$^{*}$ \\
 & TransUNet & 0.630 & \mbox{-0.152 {\tiny [-0.175, -0.127]}} & 0.97 (96\%) & \textbf{\boldmath$3.3 \times 10^{-12}\,$}$^{*}$ \\
 & StrainNet & 0.574 & \mbox{-0.104 {\tiny [-0.126, -0.080]}} & 0.84 (88\%) & \textbf{\boldmath$4.1 \times 10^{-8}\,$}$^{*}$ \\
 & Flow-matching & 0.570 & \mbox{-0.088 {\tiny [-0.118, -0.061]}} & 0.81 (86\%) & \textbf{\boldmath$1.7 \times 10^{-7}\,$}$^{*}$ \\
 & ControlNet & 0.577 & \mbox{-0.086 {\tiny [-0.108, -0.065]}} & 0.91 (88\%) & \textbf{\boldmath$1.0 \times 10^{-9}\,$}$^{*}$ \\
 & LaMoD & 0.535 & \mbox{-0.063 {\tiny [-0.088, -0.036]}} & 0.63 (80\%) & \textbf{\boldmath$1.1 \times 10^{-4}\,$}$^{*}$ \\
\midrule
\multirow{6}{*}{\makecell[l]{ES nEPE\\[2pt]\textbf{GenStrain: 0.402}}} & TLRN & 0.625 & \mbox{-0.229 {\tiny [-0.268, -0.190]}} & 0.92 (96\%) & \textbf{\boldmath$3.5 \times 10^{-10}\,$}$^{*}$ \\
 & TransUNet & 0.548 & \mbox{-0.145 {\tiny [-0.177, -0.119]}} & 0.93 (90\%) & \textbf{\boldmath$3.1 \times 10^{-10}\,$}$^{*}$ \\
 & StrainNet & 0.514 & \mbox{-0.124 {\tiny [-0.142, -0.101]}} & 0.88 (88\%) & \textbf{\boldmath$6.5 \times 10^{-9}\,$}$^{*}$ \\
 & Flow-matching & 0.478 & \mbox{-0.077 {\tiny [-0.107, -0.047]}} & 0.69 (80\%) & \textbf{\boldmath$2.0 \times 10^{-5}\,$}$^{*}$ \\
 & ControlNet & 0.510 & \mbox{-0.094 {\tiny [-0.122, -0.069]}} & 0.94 (88\%) & \textbf{\boldmath$1.0 \times 10^{-10}\,$}$^{*}$ \\
 & LaMoD & 0.442 & \mbox{-0.051 {\tiny [-0.077, -0.023]}} & 0.51 (74\%) & \textbf{\boldmath$1.5 \times 10^{-3}\,$}$^{*}$ \\
\bottomrule
\end{tabulary*}
\end{table}

\RP{Tab.~\ref{tab:supp_strain_dense} and Tab.~\ref{tab:supp_strain_cine} provide the statistical analysis of the strain MAE results corresponding to Tab.~\ref{tab:strain_error_metrics_media} in the main paper, comparing GenStrain with the baseline methods on DENSE and cine inputs, respectively.}
\begin{table}[!h]
\RPblue
\centering
\scriptsize
\setlength{\tabcolsep}{3pt}
\caption{Statistical results for DENSE MAE corresponding to Tab.~\ref{tab:strain_error_metrics_media} in the main paper. An $*$ indicates a baseline with significantly higher error than GenStrain after Holm correction across 52 comparisons.}
\label{tab:supp_strain_dense}
\begin{tabulary*}{\linewidth}{@{}LLCcCC@{}}
\toprule
 & & & \multicolumn{2}{c}{\makecell{\textbf{Paired difference}\\\textbf{(GenStrain $-$ baseline)}}} & \\
\cmidrule(lr){4-5}
\textbf{Metric} & \textbf{Baseline} & \makecell{\textbf{Baseline}\\\textbf{(mean)}} & \makecell{\textbf{HL estimate}\\\textbf{[95\% CI]}} & \makecell{\textbf{Effect size}\\\textbf{(win rate)}} & \makecell{\textbf{Holm}\\$p$} \\[1pt]
\midrule
\multirow{6}{*}{\makecell[l]{Whole-slice\\MAE (\%)\\[2pt]\textbf{GenStrain: 2.39}}} & TLRN & 5.95 & \mbox{-3.38 {\tiny [-4.14, -2.50]}} & 0.92 (88\%) & \textbf{\boldmath$3.1 \times 10^{-9}\,$}$^{*}$ \\
 & TransUNet & 3.85 & \mbox{-1.65 {\tiny [-2.15, -1.00]}} & 0.70 (82\%) & \textbf{\boldmath$9.5 \times 10^{-5}\,$}$^{*}$ \\
 & StrainNet & 3.01 & \mbox{-0.69 {\tiny [-1.08, -0.27]}} & 0.48 (68\%) & \textbf{0.023}$^{*}$ \\
 & Flow-matching & 2.96 & \mbox{-0.52 {\tiny [-0.80, -0.24]}} & 0.55 (74\%) & \textbf{\boldmath$5.6 \times 10^{-3}\,$}$^{*}$ \\
 & ControlNet & 2.89 & \mbox{-0.43 {\tiny [-0.77, -0.16]}} & 0.47 (66\%) & \textbf{0.023}$^{*}$ \\
 & LaMoD & 2.80 & \mbox{-0.45 {\tiny [-0.76, -0.16]}} & 0.47 (72\%) & \textbf{0.023}$^{*}$ \\
\midrule
\multirow{6}{*}{\makecell[l]{Whole-slice\\ES MAE (\%)\\[2pt]\textbf{GenStrain: 3.15}}} & TLRN & 6.52 & \mbox{-3.55 {\tiny [-4.57, -2.32]}} & 0.77 (84\%) & \textbf{\boldmath$1.0 \times 10^{-5}\,$}$^{*}$ \\
 & TransUNet & 5.65 & \mbox{-2.76 {\tiny [-3.71, -1.67]}} & 0.67 (84\%) & \textbf{\boldmath$2.9 \times 10^{-4}\,$}$^{*}$ \\
 & StrainNet & 4.63 & \mbox{-1.66 {\tiny [-2.41, -0.84]}} & 0.57 (72\%) & \textbf{\boldmath$3.9 \times 10^{-3}\,$}$^{*}$ \\
 & Flow-matching & 3.93 & \mbox{-0.83 {\tiny [-1.31, -0.33]}} & 0.47 (70\%) & \textbf{0.023}$^{*}$ \\
 & ControlNet & 4.39 & \mbox{-1.19 {\tiny [-1.79, -0.65]}} & 0.60 (78\%) & \textbf{\boldmath$1.8 \times 10^{-3}\,$}$^{*}$ \\
 & LaMoD & 4.09 & \mbox{-0.95 {\tiny [-1.57, -0.44]}} & 0.53 (72\%) & \textbf{\boldmath$9.1 \times 10^{-3}\,$}$^{*}$ \\
\midrule
\multirow{6}{*}{\makecell[l]{Segmental\\MAE (\%)\\[2pt]\textbf{GenStrain: 3.04}}} & TLRN & 6.97 & \mbox{-3.69 {\tiny [-4.41, -2.98]}} & 0.97 (92\%) & \textbf{\boldmath$5.0 \times 10^{-11}\,$}$^{*}$ \\
 & TransUNet & 4.77 & \mbox{-1.79 {\tiny [-2.17, -1.39]}} & 0.92 (92\%) & \textbf{\boldmath$2.6 \times 10^{-9}\,$}$^{*}$ \\
 & StrainNet & 3.87 & \mbox{-0.87 {\tiny [-1.17, -0.54]}} & 0.71 (78\%) & \textbf{\boldmath$6.9 \times 10^{-5}\,$}$^{*}$ \\
 & Flow-matching & 4.01 & \mbox{-0.88 {\tiny [-1.12, -0.66]}} & 0.82 (88\%) & \textbf{\boldmath$8.0 \times 10^{-7}\,$}$^{*}$ \\
 & ControlNet & 3.80 & \mbox{-0.70 {\tiny [-0.93, -0.50]}} & 0.84 (86\%) & \textbf{\boldmath$3.3 \times 10^{-7}\,$}$^{*}$ \\
 & LaMoD & 3.39 & \mbox{-0.39 {\tiny [-0.68, -0.08]}} & 0.41 (66\%) & \textbf{0.023}$^{*}$ \\
\midrule
\multirow{6}{*}{\makecell[l]{Segmental\\ES MAE (\%)\\[2pt]\textbf{GenStrain: 4.12}}} & TLRN & 8.33 & \mbox{-4.21 {\tiny [-5.07, -3.23]}} & 0.92 (90\%) & \textbf{\boldmath$2.9 \times 10^{-9}\,$}$^{*}$ \\
 & TransUNet & 7.26 & \mbox{-3.14 {\tiny [-3.87, -2.40]}} & 0.92 (94\%) & \textbf{\boldmath$3.8 \times 10^{-9}\,$}$^{*}$ \\
 & StrainNet & 5.79 & \mbox{-1.75 {\tiny [-2.29, -1.17]}} & 0.75 (84\%) & \textbf{\boldmath$2.2 \times 10^{-5}\,$}$^{*}$ \\
 & Flow-matching & 5.33 & \mbox{-1.23 {\tiny [-1.60, -0.81]}} & 0.71 (82\%) & \textbf{\boldmath$6.9 \times 10^{-5}\,$}$^{*}$ \\
 & ControlNet & 5.81 & \mbox{-1.53 {\tiny [-1.92, -1.15]}} & 0.91 (92\%) & \textbf{\boldmath$7.4 \times 10^{-9}\,$}$^{*}$ \\
 & LaMoD & 4.73 & \mbox{-0.64 {\tiny [-1.23, -0.10]}} & 0.36 (60\%) & \textbf{0.027}$^{*}$ \\
\bottomrule
\end{tabulary*}
\end{table}

\begin{table}[htbp]
\RPblue
\centering
\scriptsize
\setlength{\tabcolsep}{3pt}
\caption{Statistical results for cine strain MAE corresponding to Tab.~\ref{tab:strain_error_metrics_media} in the main paper. An $*$ indicates a baseline with significantly higher error than GenStrain after Holm correction across 52 comparisons.}
\label{tab:supp_strain_cine}
\begin{tabulary*}{\linewidth}{@{}LLCcCC@{}}
\toprule
 & & & \multicolumn{2}{c}{\makecell{\textbf{Paired difference}\\\textbf{(GenStrain $-$ baseline)}}} & \\
\cmidrule(lr){4-5}
\textbf{Metric} & \textbf{Baseline} & \makecell{\textbf{Baseline}\\\textbf{(mean)}} & \makecell{\textbf{HL estimate}\\\textbf{[95\% CI]}} & \makecell{\textbf{Effect size}\\\textbf{(win rate)}} & \makecell{\textbf{Holm}\\$p$} \\[1pt]
\midrule
\multirow{7}{*}{\makecell[l]{Whole-slice\\MAE (\%)\\[2pt]\textbf{GenStrain: 2.91}}} & DeepStrain & 4.60 & \mbox{-1.67 {\tiny [-2.24, -1.08]}} & 0.82 (80\%) & \textbf{\boldmath$2.0 \times 10^{-5}\,$}$^{*}$ \\
 & TLRN & 5.21 & \mbox{-2.35 {\tiny [-2.80, -1.93]}} & 0.98 (95\%) & \textbf{\boldmath$1.7 \times 10^{-9}\,$}$^{*}$ \\
 & TransUNet & 5.28 & \mbox{-2.48 {\tiny [-2.95, -1.90]}} & 0.95 (90\%) & \textbf{\boldmath$2.2 \times 10^{-8}\,$}$^{*}$ \\
 & StrainNet & 3.74 & \mbox{-0.84 {\tiny [-1.18, -0.52]}} & 0.74 (82\%) & \textbf{\boldmath$2.4 \times 10^{-4}\,$}$^{*}$ \\
 & Flow-matching & 3.92 & \mbox{-0.97 {\tiny [-1.47, -0.50]}} & 0.65 (75\%) & \textbf{\boldmath$2.9 \times 10^{-3}\,$}$^{*}$ \\
 & ControlNet & 4.51 & \mbox{-1.61 {\tiny [-2.05, -1.09]}} & 0.88 (82\%) & \textbf{\boldmath$1.8 \times 10^{-6}\,$}$^{*}$ \\
 & LaMoD & 3.60 & \mbox{-0.69 {\tiny [-1.04, -0.32]}} & 0.65 (75\%) & \textbf{\boldmath$2.9 \times 10^{-3}\,$}$^{*}$ \\
\midrule
\multirow{7}{*}{\makecell[l]{Whole-slice\\ES MAE (\%)\\[2pt]\textbf{GenStrain: 2.85}}} & DeepStrain & 4.81 & \mbox{-1.99 {\tiny [-3.03, -0.56]}} & 0.52 (68\%) & \textbf{0.023}$^{*}$ \\
 & TLRN & 5.79 & \mbox{-3.12 {\tiny [-3.81, -2.33]}} & 0.92 (88\%) & \textbf{\boldmath$2.1 \times 10^{-7}\,$}$^{*}$ \\
 & TransUNet & 7.65 & \mbox{-5.10 {\tiny [-6.08, -3.84]}} & 0.94 (88\%) & \textbf{\boldmath$6.2 \times 10^{-8}\,$}$^{*}$ \\
 & StrainNet & 5.16 & \mbox{-2.43 {\tiny [-3.20, -1.59]}} & 0.81 (82\%) & \textbf{\boldmath$2.5 \times 10^{-5}\,$}$^{*}$ \\
 & Flow-matching & 5.02 & \mbox{-2.13 {\tiny [-3.50, -0.88]}} & 0.57 (68\%) & \textbf{0.011}$^{*}$ \\
 & ControlNet & 5.94 & \mbox{-3.11 {\tiny [-3.97, -2.19]}} & 0.89 (85\%) & \textbf{\boldmath$1.1 \times 10^{-6}\,$}$^{*}$ \\
 & LaMoD & 4.70 & \mbox{-1.89 {\tiny [-2.63, -1.11]}} & 0.71 (75\%) & \textbf{\boldmath$6.4 \times 10^{-4}\,$}$^{*}$ \\
\midrule
\multirow{7}{*}{\makecell[l]{Segmental\\MAE (\%)\\[2pt]\textbf{GenStrain: 3.73}}} & DeepStrain & 5.35 & \mbox{-1.57 {\tiny [-2.07, -1.08]}} & 0.88 (85\%) & \textbf{\boldmath$1.9 \times 10^{-6}\,$}$^{*}$ \\
 & TLRN & 5.99 & \mbox{-2.34 {\tiny [-2.69, -1.94]}} & 0.99 (98\%) & \textbf{\boldmath$4.5 \times 10^{-10}\,$}$^{*}$ \\
 & TransUNet & 5.99 & \mbox{-2.28 {\tiny [-2.66, -1.88]}} & 0.99 (95\%) & \textbf{\boldmath$6.2 \times 10^{-10}\,$}$^{*}$ \\
 & StrainNet & 4.58 & \mbox{-0.84 {\tiny [-1.10, -0.62]}} & 0.85 (85\%) & \textbf{\boldmath$7.1 \times 10^{-6}\,$}$^{*}$ \\
 & Flow-matching & 5.42 & \mbox{-1.59 {\tiny [-2.05, -1.16]}} & 0.92 (90\%) & \textbf{\boldmath$1.6 \times 10^{-7}\,$}$^{*}$ \\
 & ControlNet & 5.08 & \mbox{-1.35 {\tiny [-1.72, -0.98]}} & 0.91 (85\%) & \textbf{\boldmath$2.7 \times 10^{-7}\,$}$^{*}$ \\
 & LaMoD & 4.26 & \mbox{-0.52 {\tiny [-0.75, -0.29]}} & 0.70 (78\%) & \textbf{\boldmath$9.3 \times 10^{-4}\,$}$^{*}$ \\
\midrule
\multirow{7}{*}{\makecell[l]{Segmental\\ES MAE (\%)\\[2pt]\textbf{GenStrain: 4.96}}} & DeepStrain & 6.52 & \mbox{-1.49 {\tiny [-2.47, -0.42]}} & 0.49 (65\%) & \textbf{0.023}$^{*}$ \\
 & TLRN & 7.96 & \mbox{-3.08 {\tiny [-3.64, -2.52]}} & 0.98 (92\%) & \textbf{\boldmath$2.1 \times 10^{-9}\,$}$^{*}$ \\
 & TransUNet & 9.39 & \mbox{-4.37 {\tiny [-5.20, -3.64]}} & 1.00 (100\%) & \textbf{\boldmath$9.3 \times 10^{-11}\,$}$^{*}$ \\
 & StrainNet & 7.01 & \mbox{-1.99 {\tiny [-2.58, -1.46]}} & 0.95 (90\%) & \textbf{\boldmath$2.2 \times 10^{-8}\,$}$^{*}$ \\
 & Flow-matching & 6.85 & \mbox{-1.75 {\tiny [-2.68, -0.84]}} & 0.63 (75\%) & \textbf{\boldmath$3.9 \times 10^{-3}\,$}$^{*}$ \\
 & ControlNet & 7.70 & \mbox{-2.72 {\tiny [-3.38, -2.11]}} & 0.98 (92\%) & \textbf{\boldmath$3.4 \times 10^{-9}\,$}$^{*}$ \\
 & LaMoD & 6.52 & \mbox{-1.55 {\tiny [-2.04, -1.01]}} & 0.85 (82\%) & \textbf{\boldmath$5.9 \times 10^{-6}\,$}$^{*}$ \\
\bottomrule
\end{tabulary*}
\end{table}




\clearpage
\bibliographystyle{elsarticle-num} 
\biboptions{numbers,sort&compress}
\bibliography{references}

@article{amzulescu2019myocardial,
  title={Myocardial strain imaging: review of general principles, validation, and sources of discrepancies},
  author={Amzulescu, Mihaela Silvia and De Craene, Mathieu and Langet, Hel{\'e}n{\`e} and Pasquet, Agnes and Vancraeynest, David and Pouleur, Anne-Catherine and Vanoverschelde, Jean-Louis and Gerber, BL},
  journal={European Heart Journal-Cardiovascular Imaging},
  volume={20},
  number={6},
  pages={605--619},
  year={2019},
  publisher={Oxford University Press}
}

@article{perperidis2005spatio,
  title={Spatio-temporal free-form registration of cardiac MR image sequences},
  author={Perperidis, Dimitrios and Mohiaddin, Raad H and Rueckert, Daniel},
  journal={Medical image analysis},
  volume={9},
  number={5},
  pages={441--456},
  year={2005},
  publisher={Elsevier}
}

@article{chadalavada2024myocardial,
  title={Myocardial strain measured by cardiac magnetic resonance predicts cardiovascular morbidity and death},
  author={Chadalavada, Sucharitha and Fung, Kenneth and Rauseo, Elisa and Lee, Aaron M and Khanji, Mohammed Y and Amir-Khalili, Alborz and Paiva, Jose and Naderi, Hafiz and Banik, Shantanu and Chirvasa, Mihaela and others},
  journal={Journal of the American College of Cardiology},
  volume={84},
  number={7},
  pages={648--659},
  year={2024},
  publisher={American College of Cardiology Foundation, Washington DC}
}

@article{kindberg2012myocardial,
  title={Myocardial strains from 3D displacement encoded magnetic resonance imaging},
  author={Kindberg, Katarina and Haraldsson, Henrik and Sigfridsson, Andreas and Engvall, Jan and Ingels Jr, Neil B and Ebbers, Tino and Karlsson, Matts},
  journal={BMC medical imaging},
  volume={12},
  number={1},
  pages={9},
  year={2012},
  publisher={Springer}
}

@article{sotiras2013deformable,
  title={Deformable medical image registration: A survey},
  author={Sotiras, Aristeidis and Davatzikos, Christos and Paragios, Nikos},
  journal={IEEE transactions on medical imaging},
  volume={32},
  number={7},
  pages={1153--1190},
  year={2013},
  publisher={IEEE}
}

@inproceedings{pan2021efficient,
  title={Efficient image registration network for non-rigid cardiac motion estimation},
  author={Pan, Jiazhen and Rueckert, Daniel and K{\"u}stner, Thomas and Hammernik, Kerstin},
  booktitle={International Workshop on Machine Learning for Medical Image Reconstruction},
  pages={14--24},
  year={2021},
  organization={Springer}
}

@inproceedings{chandrashekara2003construction,
  title={Construction of a statistical model for cardiac motion analysis using nonrigid image registration},
  author={Chandrashekara, Raghavendra and Rao, Anil and Sanchez-Ortiz, Gerardo Ivar and Mohiaddin, Raad H and Rueckert, Daniel},
  booktitle={Biennial International Conference on Information Processing in Medical Imaging},
  pages={599--610},
  year={2003},
  organization={Springer}
}

@article{smiseth2016myocardial,
  title={Myocardial strain imaging: how useful is it in clinical decision making?},
  author={Smiseth, Otto A and Torp, Hans and Opdahl, Anders and Haugaa, Kristina H and Urheim, Stig},
  journal={European heart journal},
  volume={37},
  number={15},
  pages={1196--1207},
  year={2016},
  publisher={Oxford University Press}
}

@article{scatteia2017strain,
  title={Strain imaging using cardiac magnetic resonance},
  author={Scatteia, Alessandra and Baritussio, Anna and Bucciarelli-Ducci, Chiara},
  journal={Heart failure reviews},
  volume={22},
  number={4},
  pages={465--476},
  year={2017},
  publisher={Springer}
}

@article{rajiah2022myocardial,
  title={Myocardial strain evaluation with cardiovascular MRI: physics, principles, and clinical applications},
  author={Rajiah, Prabhakar Shantha and Kalisz, Kevin and Broncano, Jordi and Goerne, Harold and Collins, Jeremy D and Fran{\c{c}}ois, Christopher J and Ibrahim, El-Sayed and Agarwal, Prachi P},
  journal={Radiographics},
  volume={42},
  number={4},
  pages={968--990},
  year={2022},
  publisher={Radiological Society of North America}
}

@article{gulsin2026cardiovascular,
  title={Cardiovascular Magnetic Resonance derived Myocardial Strain: Applications, Acquisition Techniques, Analysis, and Reporting. A Scientific Statement of the Society for Cardiovascular Magnetic Resonance, Radiological Society of North America, and European Association of Cardiovascular Imaging of the ESC},
  author={Gulsin, Gaurav S and Ambale-Venkatesh, Bharath and Axel, Leon and Bluemke, David A and Captur, Gabriella and Cavalcante, Jo{\~a}o L and Ennis, Daniel B and Epstein, Frederick H and Gerber, Bernhard and Hanneman, Kate and others},
  journal={Radiology},
  volume={321},
  number={1},
  pages={e261385},
  year={2026},
  publisher={Radiological Society of North America}
}

@article{lu2022dpm,
  title={Dpm-solver: A fast ode solver for diffusion probabilistic model sampling in around 10 steps},
  author={Lu, Cheng and Zhou, Yuhao and Bao, Fan and Chen, Jianfei and Li, Chongxuan and Zhu, Jun},
  journal={Advances in neural information processing systems},
  volume={35},
  pages={5775--5787},
  year={2022}
}

@article{ashburner2007fast,
  title={A fast diffeomorphic image registration algorithm},
  author={Ashburner, John},
  journal={Neuroimage},
  volume={38},
  number={1},
  pages={95--113},
  year={2007},
  publisher={Elsevier}
}

@article{beg2005computing,
  title={Computing large deformation metric mappings via geodesic flows of diffeomorphisms},
  author={Beg, M Faisal and Miller, Michael I and Trouv{\'e}, Alain and Younes, Laurent},
  journal={International journal of computer vision},
  volume={61},
  pages={139--157},
  year={2005},
  publisher={Springer}
}

@article{balakrishnan2019voxelmorph,
  title={VoxelMorph: a learning framework for deformable medical image registration},
  author={Balakrishnan, Guha and Zhao, Amy and Sabuncu, Mert R and Guttag, John and Dalca, Adrian V},
  journal={IEEE transactions on medical imaging},
  volume={38},
  number={8},
  pages={1788--1800},
  year={2019},
  publisher={IEEE}
}

@inproceedings{qin2020biomechanics,
  title={Biomechanics-informed neural networks for myocardial motion tracking in MRI},
  author={Qin, Chen and Wang, Shuo and Chen, Chen and Qiu, Huaqi and Bai, Wenjia and Rueckert, Daniel},
  booktitle={Medical Image Computing and Computer Assisted Intervention--MICCAI 2020: 23rd International Conference, Lima, Peru, October 4--8, 2020, Proceedings, Part III 23},
  pages={296--306},
  year={2020},
  organization={Springer}
}

@article{wang2023strainnet,
  title={StrainNet: Improved Myocardial Strain Analysis of Cine MRI by Deep Learning from DENSE},
  author={Wang, Yu and Sun, Changyu and Ghadimi, Sona and Auger, Daniel C and Croisille, Pierre and Viallon, Magalie and Mangion, Kenneth and Berry, Colin and Haggerty, Christopher M and Jing, Linyuan and others},
  journal={Radiology: Cardiothoracic Imaging},
  volume={5},
  number={3},
  pages={e220196},
  year={2023},
  publisher={Radiological Society of North America}
}

@inproceedings{xing2021deep,
  title={Deep Networks to Automatically Detect Late-activating Regions of the Heart},
  author={Xing, Jiarui and Ghadimi, Sona and Abdi, Mohamad and Bilchick, Kenneth C and Epstein, Frederick H and Zhang, Miaomiao},
  booktitle={2021 IEEE 18th International Symposium on Biomedical Imaging (ISBI)},
  pages={1902--1906},
  year={2021},
  organization={IEEE}
}

@article{american2002standardized,
  title={Standardized myocardial segmentation and nomenclature for tomographic imaging of the heart: a statement for healthcare professionals from the Cardiac Imaging Committee of the Council on Clinical Cardiology of the American Heart Association},
  author={American Heart Association Writing Group on Myocardial Segmentation and Registration for Cardiac Imaging: and Cerqueira, Manuel D and Weissman, Neil J and Dilsizian, Vasken and Jacobs, Alice K and Kaul, Sanjiv and Laskey, Warren K and Pennell, Dudley J and Rumberger, John A and Ryan, Thomas and others},
  journal={Circulation},
  volume={105},
  number={4},
  pages={539--542},
  year={2002},
  publisher={Am Heart Assoc}
}

@article{miller2006geodesic,
  title={Geodesic shooting for computational anatomy},
  author={Miller, Michael I and Trouv{\'e}, Alain and Younes, Laurent},
  journal={Journal of mathematical imaging and vision},
  volume={24},
  number={2},
  pages={209--228},
  year={2006},
  publisher={Springer}
}

@article{ho2020denoising,
  title={Denoising diffusion probabilistic models},
  author={Ho, Jonathan and Jain, Ajay and Abbeel, Pieter},
  journal={Advances in Neural Information Processing Systems},
  volume={33},
  pages={6840--6851},
  year={2020}
}

@article{vialard2012,
  title={Diffeomorphic 3D image registration via geodesic shooting using an efficient adjoint calculation},
  author={Vialard, Fran{\c{c}}ois X and Risser, Laurent and Rueckert, Daniel and Cotter, Colin J},
  journal={International Journal of Computer Vision},
  volume={97},
  number={2},
  pages={229--241},
  year={2012},
  publisher={Springer}
}

@article{chen2022transmorph,
  title={Transmorph: Transformer for unsupervised medical image registration},
  author={Chen, Junyu and Frey, Eric C and He, Yufan and Segars, William P and Li, Ye and Du, Yong},
  journal={Medical image analysis},
  volume={82},
  pages={102615},
  year={2022},
  publisher={Elsevier}
}

@article{miller2004computational,
  title={Computational anatomy: shape, growth, and atrophy comparison via diffeomorphisms},
  author={Miller, Michael I},
  journal={NeuroImage},
  volume={23},
  pages={S19--S33},
  year={2004},
  publisher={Elsevier}
}

@inproceedings{arsigny2006log,
  title={A log-euclidean framework for statistics on diffeomorphisms},
  author={Arsigny, Vincent and Commowick, Olivier and Pennec, Xavier and Ayache, Nicholas},
  booktitle={International Conference on Medical Image Computing and Computer-Assisted Intervention},
  pages={924--931},
  year={2006},
  organization={Springer}
}

@article{chen2021transunet,
  title={Transunet: Transformers make strong encoders for medical image segmentation},
  author={Chen, Jieneng and Lu, Yongyi and Yu, Qihang and Luo, Xiangde and Adeli, Ehsan and Wang, Yan and Lu, Le and Yuille, Alan L and Zhou, Yuyin},
  journal={arXiv preprint arXiv:2102.04306},
  year={2021}
}

@inproceedings{rueckert2006diffeomorphic,
  title={Diffeomorphic registration using B-splines},
  author={Rueckert, Daniel and Aljabar, Paul and Heckemann, Rolf A and Hajnal, Joseph V and Hammers, Alexander},
  booktitle={International Conference on Medical Image Computing and Computer-Assisted Intervention},
  pages={702--709},
  year={2006},
  organization={Springer}
}

@inproceedings{arnold1966,
  title={Sur la g{\'e}om{\'e}trie diff{\'e}rentielle des groupes de Lie de dimension infinie et ses applications {\`a} l'hydrodynamique des fluides parfaits},
  author={Arnold, Vladimir},
  booktitle={Annales de l'institut Fourier},
  volume={16},
  number={1},
  pages={319--361},
  year={1966}
}

@article{cao2018comparison,
  title={A comparison of both DENSE and feature tracking techniques with tagging for the cardiovascular magnetic resonance assessment of myocardial strain},
  author={Cao, J Jane and Ngai, Nora and Duncanson, Lynette and Cheng, Joshua and Gliganic, Kathleen and Chen, Qizhi},
  journal={Journal of Cardiovascular Magnetic Resonance},
  volume={20},
  number={1},
  pages={26},
  year={2018},
  publisher={Springer}
}

@article{bucius2020comparison,
  title={Comparison of Feature Tracking, Fast-Senc, and Myocardial Tagging for Global and Semental Left Ventricular Strain},
  author={Bucius, Paulius and Erley, Jennifer and Tanacli, Radu and Zieschang, Victoria and Giusca, Sorin and Korosoglou, Grigorious and Steen, Henning and Stehning, Christian and Pieske, Burkert and Pieske-Kraigher, Elisabeth and others},
  journal={ESC heart failure},
  volume={7},
  number={2},
  pages={523--532},
  year={2020},
  publisher={Oxford University Press}
}

@article{morales2021deepstrain,
  title={DeepStrain: a deep learning workflow for the automated characterization of cardiac mechanics},
  author={Morales, Manuel A and Van den Boomen, Maaike and Nguyen, Christopher and Kalpathy-Cramer, Jayashree and Rosen, Bruce R and Stultz, Collin M and Izquierdo-Garcia, David and Catana, Ciprian},
  journal={Frontiers in Cardiovascular Medicine},
  volume={8},
  pages={730316},
  year={2021},
  publisher={Frontiers}
}

@article{qin2023generative,
  title={Generative myocardial motion tracking via latent space exploration with biomechanics-informed prior},
  author={Qin, Chen and Wang, Shuo and Chen, Chen and Bai, Wenjia and Rueckert, Daniel},
  journal={Medical Image Analysis},
  volume={83},
  pages={102682},
  year={2023},
  publisher={Elsevier}
}

@article{zhong2010imaging,
  title={Imaging three-dimensional myocardial mechanics using navigator-gated volumetric spiral cine DENSE MRI},
  author={Zhong, Xiaodong and Spottiswoode, Bruce S and Meyer, Craig H and Kramer, Christopher M and Epstein, Frederick H},
  journal={Magnetic resonance in medicine},
  volume={64},
  number={4},
  pages={1089--1097},
  year={2010},
  publisher={Wiley Online Library}
}

@article{schuster2016cardiovascular,
  title={Cardiovascular magnetic resonance myocardial feature tracking: concepts and clinical applications},
  author={Schuster, Andreas and Hor, Kan N and Kowallick, Johannes T and Beerbaum, Philipp and Kutty, Shelby},
  journal={Circulation: Cardiovascular Imaging},
  volume={9},
  number={4},
  pages={e004077},
  year={2016},
  publisher={Am Heart Assoc}
}

@article{xu2022state,
  title={State-of-the-art myocardial strain by CMR feature tracking: clinical applications and future perspectives},
  author={Xu, Jing and Yang, Wenjing and Zhao, Shihua and Lu, Minjie},
  journal={European radiology},
  volume={32},
  number={8},
  pages={5424--5435},
  year={2022},
  publisher={Springer}
}

@article{morais2017cardiovascular,
  title={Cardiovascular magnetic resonance myocardial feature tracking using a non-rigid, elastic image registration algorithm: assessment of variability in a real-life clinical setting},
  author={Morais, Pedro and Marchi, Alberto and Bogaert, Julie A and Dresselaers, Tom and Heyde, Brecht and D’hooge, Jan and Bogaert, Jan},
  journal={Journal of Cardiovascular Magnetic Resonance},
  volume={19},
  number={1},
  pages={24},
  year={2017},
  publisher={Springer}
}

@article{wehner2018comparison,
  title={Comparison of left ventricular strains and torsion derived from feature tracking and DENSE CMR},
  author={Wehner, Gregory J and Jing, Linyuan and Haggerty, Christopher M and Suever, Jonathan D and Chen, Jing and Hamlet, Sean M and Feindt, Jared A and Dimitri Mojsejenko, W and Fogel, Mark A and Fornwalt, Brandon K},
  journal={Journal of Cardiovascular Magnetic Resonance},
  volume={20},
  number={1},
  pages={63},
  year={2018},
  publisher={Springer}
}

@article{markl2011comprehensive,
  title={Comprehensive 4D velocity mapping of the heart and great vessels by cardiovascular magnetic resonance},
  author={Markl, Michael and Kilner, Philip J and Ebbers, Tino},
  journal={Journal of Cardiovascular Magnetic Resonance},
  volume={13},
  number={1},
  pages={7},
  year={2011},
  publisher={Elsevier}
}

@article{kim2004myocardial,
  title={Myocardial tissue tracking with two-dimensional cine displacement-encoded MR imaging: development and initial evaluation},
  author={Kim, Daniel and Gilson, Wesley D and Kramer, Christopher M and Epstein, Frederick H},
  journal={Radiology},
  volume={230},
  number={3},
  pages={862--871},
  year={2004},
  publisher={Radiological Society of North America}
}

@article{fischer2020feature,
  title={Feature tracking myocardial strain incrementally improves prognostication in myocarditis beyond traditional CMR imaging features},
  author={Fischer, Kady and Obrist, Sarah J and Erne, Sophie A and Stark, Anselm W and Marggraf, Maximilian and Kaneko, Kyoichi and Guensch, Dominik P and Huber, Adrian T and Greulich, Simon and Aghayev, Ayaz and others},
  journal={Cardiovascular Imaging},
  volume={13},
  number={9},
  pages={1891--1901},
  year={2020},
  publisher={American College of Cardiology Foundation, Washington DC}
}

@inproceedings{xing2024lamod,
  title={LaMoD: Latent Motion Diffusion Model for Myocardial Strain Generation},
  author={Xing, Jiarui and Jayakumar, Nivetha and Wu, Nian and Wang, Yu and Epstein, Frederick H and Zhang, Miaomiao},
  booktitle={International Workshop on Shape in Medical Imaging},
  pages={164--177},
  year={2024},
  organization={Springer}
}

@inproceedings{wu2024tlrn,
  title={TLRN: Temporal Latent Residual Networks for Large Deformation Image Registration},
  author={Wu, Nian and Xing, Jiarui and Zhang, Miaomiao},
  booktitle={International Conference on Medical Image Computing and Computer-Assisted Intervention},
  pages={728--738},
  year={2024},
  organization={Springer}
}

@article{ho2022video,
  title={Video diffusion models},
  author={Ho, Jonathan and Salimans, Tim and Gritsenko, Alexey and Chan, William and Norouzi, Mohammad and Fleet, David J},
  journal={Advances in neural information processing systems},
  volume={35},
  pages={8633--8646},
  year={2022}
}

@inproceedings{lei2025dense,
  title={DENSE-Guided Deep Motion Networks Accounted by Large Rotations to Improve Myocardial Strain Analysis from Routine Cine MRI},
  author={Lei, Pengcheng and Xing, Jiarui and Fang, Faming and Epstein, Frederick H and Zhang, Miaomiao},
  booktitle={2025 IEEE 22nd International Symposium on Biomedical Imaging (ISBI)},
  pages={1--5},
  year={2025},
  organization={IEEE}
}

@article{spottiswoode2007tracking,
  title={Tracking myocardial motion from cine DENSE images using spatiotemporal phase unwrapping and temporal fitting},
  author={Spottiswoode, Bruce S and Zhong, Xiaodong and Hess, Aaron T and Kramer, CM and Meintjes, Ernesta M and Mayosi, Bongani M and Epstein, Frederick H},
  journal={IEEE transactions on medical imaging},
  volume={26},
  number={1},
  pages={15--30},
  year={2007},
  publisher={IEEE}
}

@article{ghadimi2023improved,
  title={Improved computation of Lagrangian tissue displacement and strain for cine DENSE MRI using a regularized spatiotemporal least squares method},
  author={Ghadimi, Sona and Abdi, Mohamad and Epstein, Frederick H},
  journal={Frontiers in Cardiovascular Medicine},
  volume={10},
  pages={1095159},
  year={2023},
  publisher={Frontiers Media SA}
}

@article{shehata2009myocardial,
  title={Myocardial tissue tagging with cardiovascular magnetic resonance},
  author={Shehata, Monda L and Cheng, Susan and Osman, Nael F and Bluemke, David A and Lima, Jo{\~a}o AC},
  journal={Journal of Cardiovascular Magnetic Resonance},
  volume={11},
  number={1},
  pages={55},
  year={2009},
  publisher={Elsevier}
}

@article{zerhouni1988human,
  title={Human heart: tagging with MR imaging--a method for noninvasive assessment of myocardial motion.},
  author={Zerhouni, Elias A and Parish, David M and Rogers, Walter J and Yang, Andrew and Shapiro, Edward P},
  journal={Radiology},
  volume={169},
  number={1},
  pages={59--63},
  year={1988}
}

@article{ferdian2020fully,
  title={Fully automated myocardial strain estimation from cardiovascular MRI--tagged images using a deep learning framework in the UK Biobank},
  author={Ferdian, Edward and Suinesiaputra, Avan and Fung, Kenneth and Aung, Nay and Lukaschuk, Elena and Barutcu, Ahmet and Maclean, Edd and Paiva, Jose and Piechnik, Stefan K and Neubauer, Stefan and others},
  journal={Radiology: Cardiothoracic Imaging},
  volume={2},
  number={1},
  pages={e190032},
  year={2020},
  publisher={Radiological Society of North America}
}

@inproceedings{lipman2023flow,
title={Flow Matching for Generative Modeling},
author={Yaron Lipman and Ricky T. Q. Chen and Heli Ben-Hamu and Maximilian Nickel and Matthew Le},
booktitle={The Eleventh International Conference on Learning Representations },
year={2023},
url={https://openreview.net/forum?id=PqvMRDCJT9t}
}

@article{vaswani2017attention,
  title={Attention is all you need},
  author={Vaswani, Ashish and Shazeer, Noam and Parmar, Niki and Uszkoreit, Jakob and Jones, Llion and Gomez, Aidan N and Kaiser, {\L}ukasz and Polosukhin, Illia},
  journal={Advances in neural information processing systems},
  volume={30},
  year={2017}
}

@article{Kingma2014AdamAM,
  title={Adam: A Method for Stochastic Optimization},
  author={Diederik P. Kingma and Jimmy Ba},
  journal={CoRR},
  year={2014},
  volume={abs/1412.6980},
  url={https://api.semanticscholar.org/CorpusID:6628106}
}

@article{song2020denoising,
  title={Denoising diffusion implicit models},
  author={Song, Jiaming and Meng, Chenlin and Ermon, Stefano},
  journal={arXiv preprint arXiv:2010.02502},
  year={2020}
}

@article{bergstra2012random,
  title={Random search for hyper-parameter optimization.},
  author={Bergstra, James and Bengio, Yoshua},
  journal={Journal of machine learning research},
  volume={13},
  number={2},
  year={2012}
}

@article{song2020score,
  title={Score-based generative modeling through stochastic differential equations},
  author={Song, Yang and Sohl-Dickstein, Jascha and Kingma, Diederik P and Kumar, Abhishek and Ermon, Stefano and Poole, Ben},
  journal={arXiv preprint arXiv:2011.13456},
  year={2020}
}

@article{shrout1979intraclass,
  title={Intraclass correlations: uses in assessing rater reliability.},
  author={Shrout, Patrick E and Fleiss, Joseph L},
  journal={Psychological bulletin},
  volume={86},
  number={2},
  pages={420},
  year={1979},
  publisher={American Psychological Association}
}

@incollection{benesty2009pearson,
  title={Pearson correlation coefficient},
  author={Benesty, Jacob and Chen, Jingdong and Huang, Yiteng and Cohen, Israel},
  booktitle={Noise reduction in speech processing},
  pages={1--4},
  year={2009},
  publisher={Springer}
}

@article{reed2002use,
  title={Use of coefficient of variation in assessing variability of quantitative assays},
  author={Reed, George F and Lynn, Freyja and Meade, Bruce D},
  journal={Clinical and Vaccine Immunology},
  volume={9},
  number={6},
  pages={1235--1239},
  year={2002},
  publisher={American Society for Microbiology}
}

@misc{zhang2023adding,
  title={Adding Conditional Control to Text-to-Image Diffusion Models}, 
  author={Lvmin Zhang and Anyi Rao and Maneesh Agrawala},
  booktitle={IEEE International Conference on Computer Vision (ICCV)},
  year={2023},
}

@article{tong2023improving,
  title={Improving and generalizing flow-based generative models with minibatch optimal transport},
  author={Tong, Alexander and Fatras, Kilian and others},
  journal={arXiv preprint arXiv:2302.00482},
  year={2023}
}

@incollection{wilcoxon1992individual,
  title={Individual comparisons by ranking methods},
  author={Wilcoxon, Frank},
  booktitle={Breakthroughs in statistics: Methodology and distribution},
  pages={196--202},
  year={1992},
  publisher={Springer}
}

@article{holm1979simple,
  title={A simple sequentially rejective multiple test procedure},
  author={Holm, Sture},
  journal={Scandinavian journal of statistics},
  pages={65--70},
  year={1979},
  publisher={JSTOR}
}

@incollection{hodges2011estimates,
  title={Estimates of location based on rank tests},
  author={Hodges Jr, Joseph L and Lehmann, Erich L},
  booktitle={Selected works of EL Lehmann},
  pages={287--300},
  year={2011},
  publisher={Springer}
}

@article{kerby2014simple,
  title={The simple difference formula: An approach to teaching nonparametric correlation},
  author={Kerby, Dave S},
  journal={Comprehensive Psychology},
  volume={3},
  pages={11--IT},
  year={2014},
  publisher={SAGE Publications Sage CA: Los Angeles, CA}
}

@article{yingchoncharoen2013normal,
  title={Normal ranges of left ventricular strain: a meta-analysis},
  author={Yingchoncharoen, Teerapat and Agarwal, Shikhar and Popovi{\'c}, Zoran B and Marwick, Thomas H},
  journal={Journal of the American Society of Echocardiography},
  volume={26},
  number={2},
  pages={185--191},
  year={2013},
  publisher={Elsevier}
}

\end{document}